\documentclass[preprint, 12pt]{elsarticle}
\usepackage{lineno}
\usepackage{hyperref}
\hypersetup{colorlinks = true}
\journal{Journal of Aerosol Science}
\biboptions{authoryear}
\usepackage{amsmath}
\usepackage{caption}
\usepackage{subcaption}
\usepackage{graphicx}
\usepackage{upgreek}
\graphicspath{ {img/} }
\usepackage{geometry}
\usepackage{array}
\usepackage{upgreek} 
\usepackage{pdflscape}
\usepackage{enumitem}
\setitemize{noitemsep,topsep=0pt,parsep=0pt,partopsep=0pt}
\setenumerate{noitemsep,topsep=0pt,parsep=0pt,partopsep=0pt}

\usepackage{setspace}
 
\usepackage[T1]{fontenc}
\usepackage{kpfonts}

\begin{document}
\begin{frontmatter}
\title{Spray characteristics from nasal spray atomization}

\author[add4]{James Van Strien}
\author[add1]{Phred Petersen}
\author[add1]{Petros Lappas}
\author[add3]{Leslie Yeo}
\author[add3]{Amgad Rezk}
\author[add1,add4]{Sara Vahaji}
\author[add1]{Kiao Inthavong\corref{cor1}}\ead{kiao.inthavong@rmit.edu.au}

\cortext[cor1]{Corresponding Author}

\address[add1]{Mechanical \& Automotive Engineering, School of Engineering, RMIT University, Bundoora, Australia}
\address[add3]{Micro/Nanophysics Research Laboratory, School of Engineering, RMIT University, Melbourne, Australia}
\address[add4]{Faculty of Sci Eng \& Built Env, School of Engineering, Deakin University, Geelong Waurn Ponds, Australia}

\begin{abstract}
\setstretch{1.0}In addition to topical delivery, nasal sprays offer an alternative drug administration route to the systemic circulation, therefore avoiding the need for painful, invasive delivery techniques. Studies on the efficacy of nasal drug delivery to date have been conflicting despite its potential. In particular computational studies, e.g. Computational Fluid Dynamics (CFD), of nasal spray deposition require realistic initial spray conditions, which are lacking in the literature. This study aims to provide benchmark experimental data for CFD inputs by characterising four different types of commercially available nasal sprays used in an ENT (Ear Nose Throat) clinic. Spray characteristics were obtained via high-speed videography, while droplet size distribution (DSD) measurements were obtained with laser diffraction. Three actuation forces were evaluated, representing the use-case of an average adult, average child and the maximum force applied to the spray device. The results demonstrated that actuation force influenced spray velocity, spray duration, DSD, and breakup length. Quantitative data of Rosin-Rammler diameter distributions, spray cone angles, dispersion angles and breakup lengths were found, assisting computational models with realistic values, thereby improving future CFD studies of nasal spray drug delivery in the nasal cavity.
\end{abstract}

\end{frontmatter}

\section{Introduction}
\label{Introduction}
Nasal drug delivery can deliver systemically acting drugs due to the highly vascularised mucosa directly connected to the bloodstream. For nasal sprays to deliver its full potential, insights into nasal spray atomization is needed to achieve target drug delivery to specific parts of the nasal cavity. For example, \cite{Foo2007} investigated three spray devices and found that spray angle and delivery technique influenced nasal deposition within the turbinate region. A wider spray plume provided less deposition efficiency but was less influenced by administration technique, while inhalation and droplet size had minimal effect, however this is likely due to the Dv50 values being greater than $20\upmu$m, as \cite{inthavong2006numerical} showed that droplets greater than $20\upmu$m exhibited high momentum leading to immediate deposition, whereas droplets $<10\upmu$m are strongly influenced by the airflow.

\cite{Suman2002} compared two aqueous spray pumps in vivo and in vitro on 10 subjects and found that while there were significant differences in the spray characteristics (spray pattern, plume geometry and droplet size); the deposition patterns from the in vivo tests were indistinguishable. These results are in line with the work of \cite{Foo2007}, who found that droplet size had minimal effect on deposition efficiency, again due to Dv50 greater than $20\upmu$m. It is also likely that the deposition efficiency in the \cite{Suman2002} study was less influenced by administration technique as the $~60^\circ$ spray plume used may be a contributing factor to the similar deposition patterns in vivo. 

\cite{Cheng2001} evaluated deposition in a cast nasal cavity model from four different nasal spray pumps. They concluded that smaller droplets and a narrower spray angle improved deposition within the turbinate region however, the study did not specify insertion angle of the spray. While the conclusion regarding spray angle would reflect the work of \cite{Foo2007}, it differs from preceding work in saying that droplet size has an affect on particle deposition within the nasal cavity. The conflicting results between studies suggests deposition is multi-factorial and merits further investigation which could be assisted by CFD. 

CFD studies include \cite{kimbell2007characterization} which found that deposition efficiency increased when the nozzle tip was inserted 1.0cm into the nostril, rather than further out. \cite{inthavong2008optimising,inthavong2011simulation} demonstrated that particle size, cone diameter at the breakup length and spray cone angle were significant factors on deposition efficacy. \cite{dong2018partitioning} found a relationship between the origin of inhaled nanoparticles and its deposition location in a nasal cavity. \cite{tong2016effects} determined the ideal delivery axis to target nasal spray deposition to the middle meatus. These CFD investigations used monodisperse particles and spray characteristics that do not necessarily represent realistic spray conditions, suggesting the need for experimental data quantifying spray characteristics to improve these computer models.

An alternative approach for spray modelling is using atomization breakup models that represent the liquid sheet breakup and produce droplets from input data. For example, the Linearized Instability Sheet atomization (LISA) model \citep{schmidt1999pressure,gavtash2014linear}, produces polydisperse droplets with initial velocities based on the primary breakup stage of atomization. Such models are efficient as they do not require substantial computational resources compared with fully resolved simulations of liquid sheet breakup. The LISA model is evaluated through instability theory of wave propagation and requires input values that characterise the atomizer and the physical atomization process within the secondary disperse phase modelling of CFD investigations. These input values are typically obtained from experimental measurements that are not widely available in the literature, particularly the dispersion angle, which currently has only one data set in the literature in nasal sprays. An objective of this study is to address this lack of available data and provide a comprehensive set of experimental data from multiple spray devices.

Experimental studies characterising nasal sprays include \cite{Inthavong2015} which found that increasing actuation force increased spray velocity, decreased droplet size and decreased spray time. \cite{Fung2013} used high-speed shadowgraphy and image analysis to characterise the behaviour of water sprayed from a Beconase Hayfever nasal spray bottle. They found four distinct stages of the development of the spray; distorted pencil, onion stage, tulip stage and fully developed stage. These stages are also found in high-pressure applications \citep{lefebvre2017atomization}. The results of \cite{Dayal2004} demonstrated that an increase in viscosity caused a linear increase in average droplet size, and an increase in actuation force lowered the average droplet size and created a narrower Droplet Size Distribution (DSD). \cite{Guo2008} found that an increase in viscosity produced a narrower plume geometry, a wider DSD and an increase in the average droplet size. An increase in actuation speed caused a narrower DSD and plume geometry while lowering the average droplet size. These experimental investigations provide an understanding of actuation force and viscosity on the spray atomization; however, only some quantitative spray parameters required for setting up the computational atomization breakup models were reported.

This study aims to provide benchmark experimental data for spray atomization models by characterising four different types of commercially available nasal sprays used in an ENT (Ear Nose Throat) clinic. Spray characteristics were obtained via high-speed videography, while droplet size distribution (DSD) measurements were obtained with laser diffraction. Three actuation forces were evaluated, representing the use-case of an average adult, average child and the maximum force applied to the spray device. The results demonstrated that actuation force influenced spray velocity, spray duration, DSD, and breakup length. Quantitative data of Rosin-Rammler diameter distributions, spray cone angles, dispersion angles and breakup lengths were found, assisting computational models with realistic values to setup the initial spray conditions, thereby improving future CFD studies of nasal spray drug delivery in the nasal cavity. 


\section{Methods and Materials}
\begin{figure}[h]
\centering\includegraphics[width=0.8\linewidth]{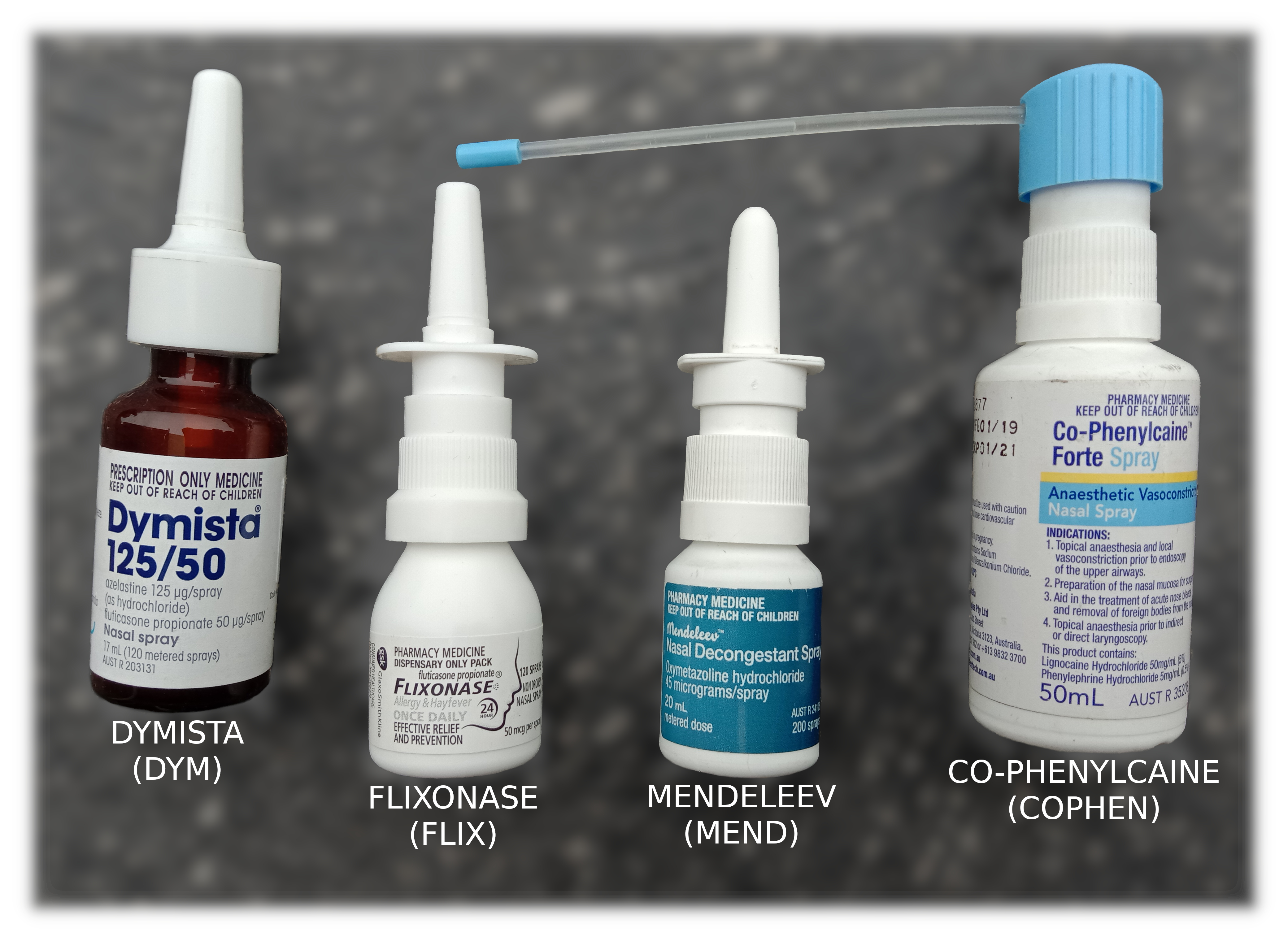}
\caption{A photo of the four nasal sprays investigated, from left: \textit{Dymista} antihistamine (DYM), \textit{Flixonase} anti-inflammatory (FLIX), \textit{Mendeleev} nasal decongestant (MEND) and \textit{Co-Phenylcaine} anaesthetic and vasoconstrictor (COPHEN). Note that COPHEN is the only nozzle with a horizontal orientation. }\label{fig:fourBottles}
\end{figure}

Four commonly used nasal spray devices used in an ENT clinic were supplied by a practising ENT clinician. The devices included: \textit{Mendeleev} nasal decongestant (MEND), \textit{Co-Phenylcaine} anaesthetic and vasoconstrictor (COPHEN), \textit{Dymista} antihistamine (DYM) and \textit{Flixonase} anti-inflammatory (FLIX), see figure \ref{fig:fourBottles}. The devices produced a spray in the vertical direction aligned with the bottle axis, except for the COPHEN device that dispensed the formulation via a horizontal tube (called a `Flexi Nozzle') which is attached to the device. The atomizer was located at the end of the horizontal tube and produced a spray perpendicular to the bottle.

\subsection{Actuation Device}
An automatic nasal spray actuator device was built for consistent and repeatable actuation, as studies have shown variability among human actuation \citep{Doughty2011,Kippax2000}. The actuator consisted of an air pressure regulator, a 4-way two position valve operated by an electric solenoid and controlled by an Arduino micro-controller. The actuator was located inside a frame that held the nasal spray bottle. The actuator moved the bottle while locking the nozzle stationary. Since a solenoid switch applies a step function (for on/off switches) the applied force over time is also a step function, and was applied for the entire spray period.

The air pressure regulator was adjusted to create the desired actuation force imposed on the nasal spray. Pressures of 0.2, 0.4 and 0.6 MPa were used. The area on the out-stroke side of the actuator piston was 157 mm$^2$, hence the approximate forces on the bottle were 31, 63 and 94 Newtons. This provided values close to the paediatric and adult mean actuation forces, as well as the maximum adult force, respectively \citep{Doughty2011}.

\subsection{High Speed Imaging}\label{subsec:HSI}
A Phantom Miro M310 high speed camera was used to record up-close video of the sprays for characterising cone angle, dispersion angle, phase duration and breakup length at the varying actuation forces. 

\begin{figure}[h]
\centering\includegraphics[width=0.8\linewidth]{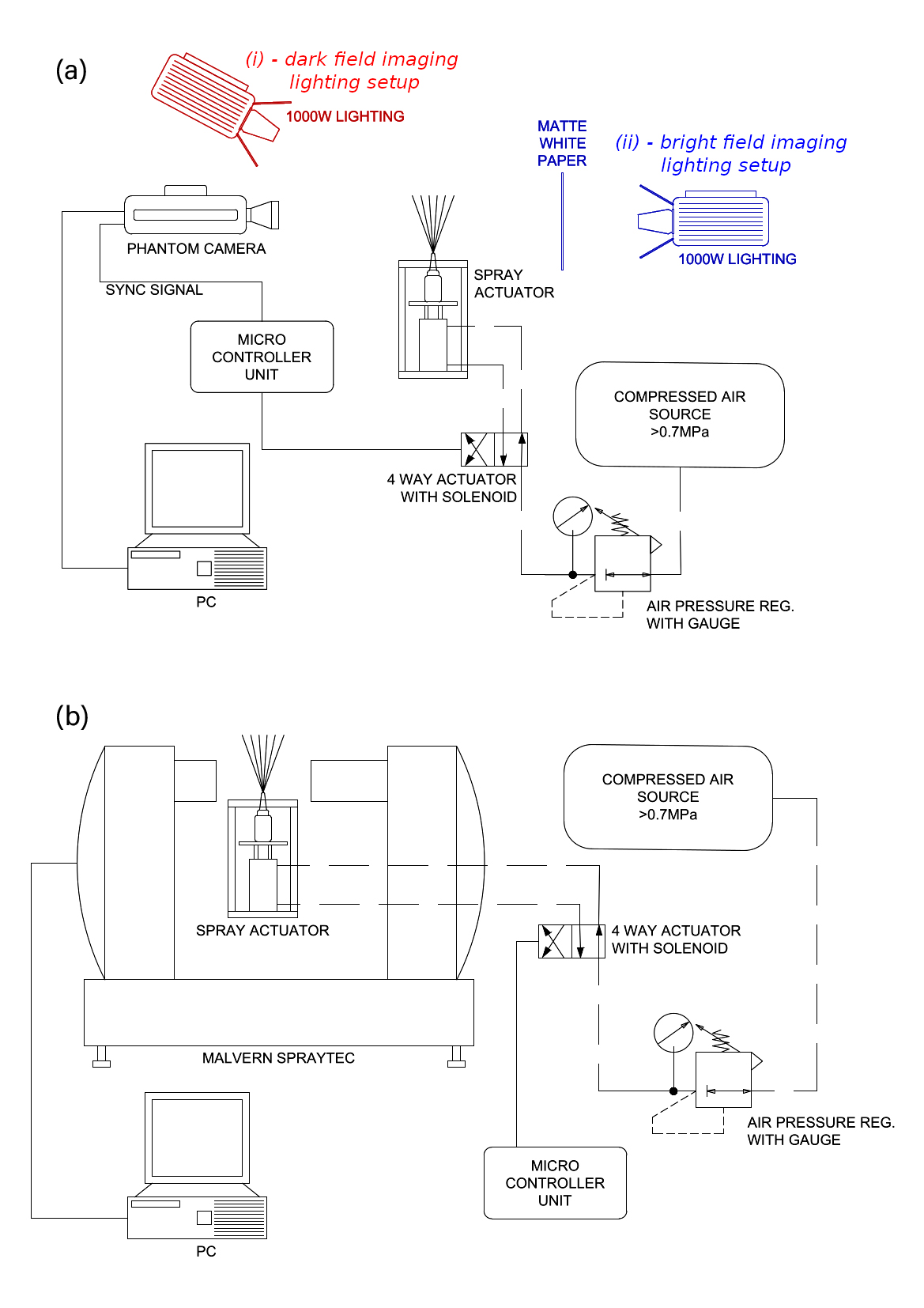}
\caption{(a) Schematic of high speed videography where (i) depicts foreground lighting of the dark field imaging in blue and (ii) shows the background lighting technique for bright field imaging in red. (b) Schematic of the experimental setup for DSD capture with the Malvern Spraytec.}\label{fig:filmingMalvernSchematic}
\end{figure}

Two illumination methods and field of views were used for the high-speed videography. Bright field illumination was used for close-up footage with a viewing pane of around 8 x 10 mm. A thin white background illuminated from behind the subject causes the background to appear white while the aerosol casts a shadow in the image, appearing black. For the wide shots with a viewing pane of around 33 x 42 mm, darkfield illumination was used. The aerosol is front lit and with enough distance to the nearest backdrop, the background of the footage is black.  Figure \ref{fig:filmingMalvernSchematic} (a) contains a schematic of this methodology. 

For each spray, the actuator was mounted to a table with clamps. Once the actuator device was securely located, the camera and lighting were set up around the actuator. The lab's HVAC system was turned off to prevent airflow from affecting the behaviour of the sprays. Each spray bottle was filmed with the original drug formulation being sprayed. For the Flixonase and Mendeleeve bottles, the drug formulation was replaced with distilled water and filmed for comparison. Each bottle was primed up to ten times by hand until a consistent resistance was felt throughout the range of motion of the pump. Each scenario was executed and filmed four times to ensure consistency of results. 

\cite{Fung2013} found large water ligaments in the spray plume during atomization resulting from residual liquid pooling in the nozzle. Their method used an absorbent media to clean the nozzle between sprays which may not remove the pooled fluid. In this study, compressed air was used to clear the nozzle tip of the spray bottle to prevent pooling between each spray. No ligaments were found during the initial stages of this footage.

\subsection{Droplet Size Distribution Measurements}
Droplet size distribution (DSD) was captured with a Malvern Spraytec\textregistered\hspace{1pt} laser diffraction system and the accompanying Spraytec 3.20 software. Figure \ref{fig:filmingMalvernSchematic} (b) provides a schematic of the setup. Each spray was primed, described in section \ref{subsec:HSI} and the room ventilation system was turned off to prevent ambient air currents affecting the spray results. The spray actuator device was mounted on an adjustable platform to achieve distances of 15, 30, and 60 mm between the laser beam and the nozzle tip. This was repeated for actuation forces of 31, 63 and 94 Newtons and each spray was repeated three times at each distance/force combination. Between each spray, compressed air was used to clear the nozzle tip of the spray bottle to prevent pooling. 

The measured DSD should be taken close to the atomization region, i.e. where the droplets form. The average breakup length has been reported as approximately 6~mm \citep{Fung2013, inthavong2014high}, and we selected 15~mm as the closest location was to allow a sufficient distance for the liquid sheet to breakup into ligaments and then droplets \citep{shrestha2020primary}. The overall distance of 15~mm is a compromise between getting as close to the primary breakup as possible while still achieving consistent results.Furthermore, the work by \cite{Dayal2004} also measured at distances of 15, 30, and 60~mm distances.

\begin{figure}[h]
\centering\includegraphics[width=0.8\linewidth]{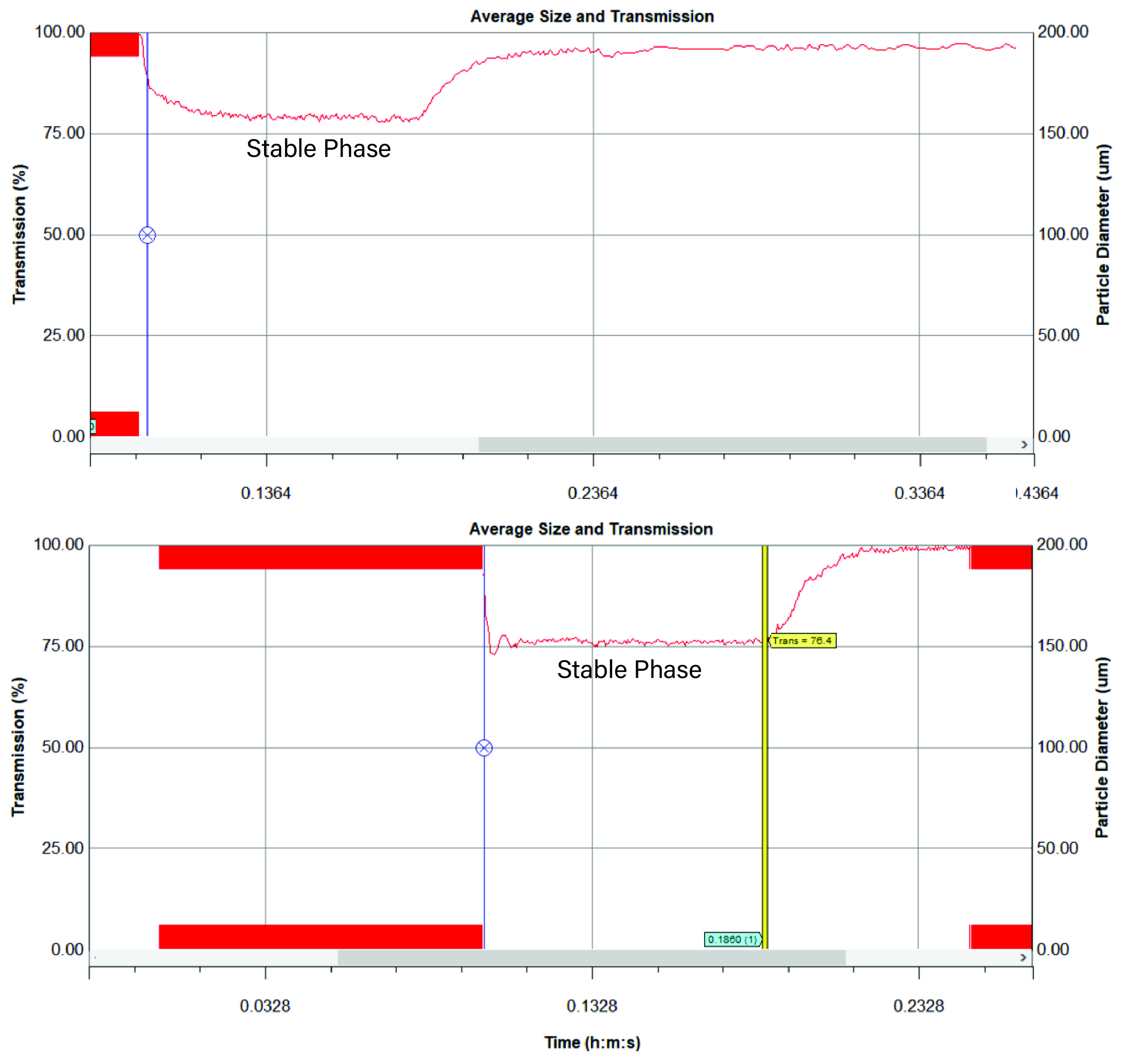}
\caption{Sample screenshots of the Malvern Spraytec transmission levels during data acquisition. The solid horizontal red bars represent no interference with the laser, and therefore no spray detected. The stable phase is labelled which is the region of minimum transmission.}
\label{fig:spraytec}
\end{figure}

Data was acquired with the Malvern Spraytec before the spray began and continued logging data beyond the spray period, thus the data is continuously logged over the entire spray period. To ensure we take the stable phase of the spray, we only retain data during the the stable phase based on the steady minimum transmission levels. Sample screenshots from the Malvern Spraytec is shown in Figure \ref{fig:spraytec}, and the stable phase where data was collected is labelled.

\subsection{Image Processing}
The close-up, bright-field high-speed imaging results were used to find the spray cone angle for each spray device. The images of each event were limited to the fully developed spray stage. Image processing was performed with Python OpenCV using Canny edge detection \citep{canny1986computational}. The script used the resulting Canny image to capture the first and last dark pixel of each pixel row, representing the outer edge of the spray cone at that frame. The script output produced two separate images; a heatmap that represents the most common pixel locations that the spray cone covered and an overlay that showed the spray cone's minimum and maximum limits during the developed spray regime. 

The spray breakup length was found through manual processing of the images from samples of the fully developed spray stage. The breakup length was defined as the distance from the nozzle tip, to the end of the solid spray cone. The data was taken from a sample of 20 frames from each filmed event by measuring the number of pixels from the nozzle tip to the end of the spray cone sheet. 

Spray penetration with time was determined by detecting the leading edge of the initial spray droplets from the wide view, dark field images. Individual images of the spray with time were image-processed where the spray penetration over the time was determined, thus provides an estimate of the averaged spray axial velocity during the initial spray development. In reality, individual droplets exhibit three dimensional velocity components, and therefore, our image processing technique captures the spray penetration velocity. As the droplets move with the spray, the spray penetration velocity represents the initial droplet axial velocity during the initial atomisation period. This measurement differs from Phase Doppler Particle Analyzer (PDPA) analysis which captures individual droplet size, and velocity components simultaneously at a single precise location.

\section{Results}
Abbreviations used in the results refer to the brand spray bottle defined as follows: COPHEN = Co-phenylcaine, DYM = Dymista, FLIX = Flixonase, MEND = Mendeleev. For the Flixonase and Mendeleev bottle, the suffix \textit{\_D} represents the drug formulation, and \textit{\_W} represents water that was used instead of the drug.

\subsection{Spray Development Stages} 
\begin{figure}[h]
\centering\includegraphics[width=0.9\linewidth]{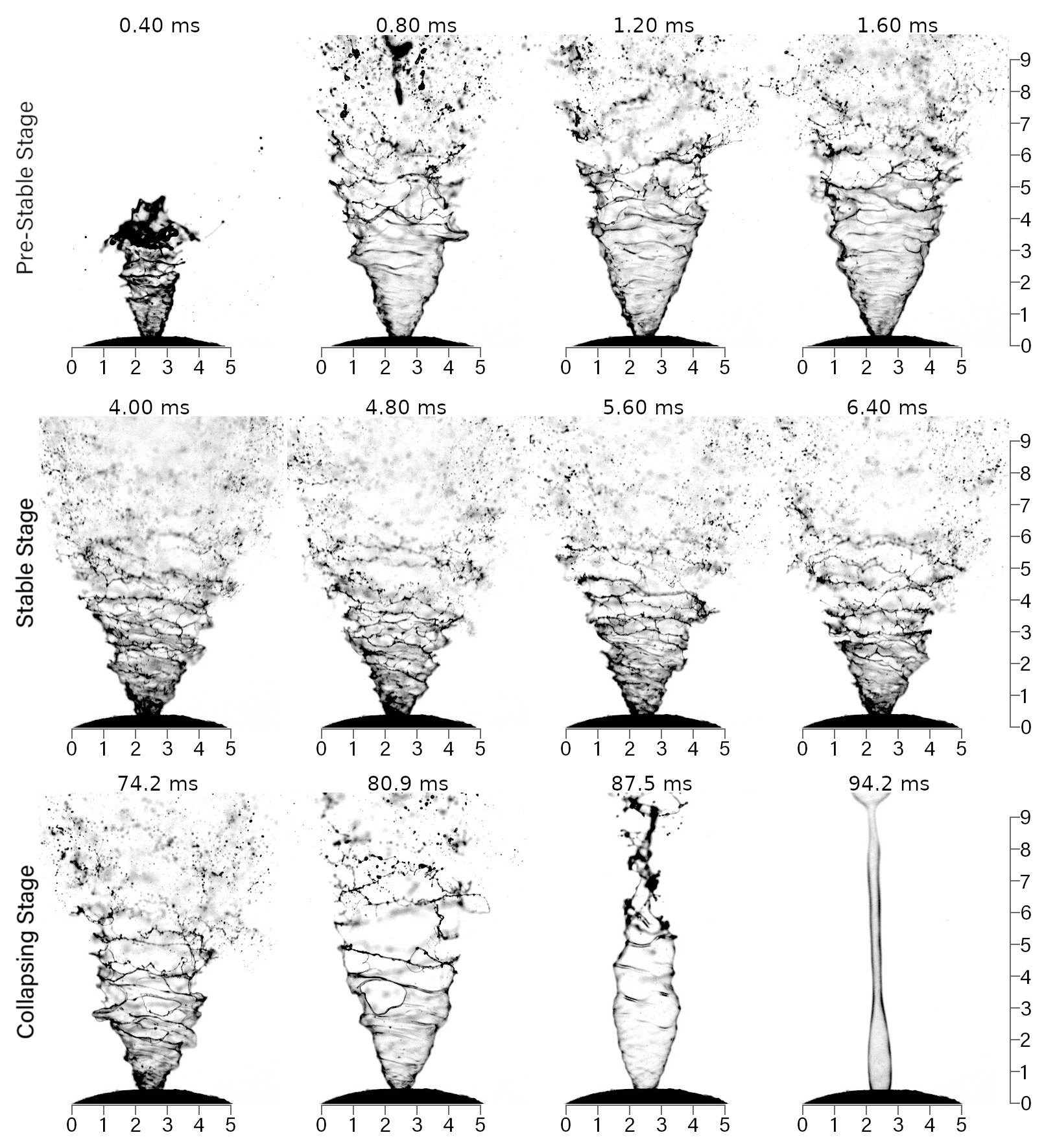}
\caption{A sample of four images from each stage of Dymista actuated at 94 N. The top row shows the pre-stable stage at a sample rate of 0.4 ms. The middle row demonstrates the stable stage at a frame rate of 0.8 ms and the bottom row shows the collapsing stage at sample rate of 3.33 ms. The axes dimensions are in millimeters (mm).}\label{fig:phaseSample}
\end{figure}

The spray plume development during atomization was categorised into three stages following \cite{Fung2013} that suggested  a developing stage, a stable developed stage, and a collapsing stage. Figure \ref{fig:phaseSample} provides a representative sample of the atomization event through eight images for each phase for the Dymista atomization actuated at 94 N. The images show a swirling liquid sheet that breaks up within 0.8 ms. At the breakup region, the process is highly erratic, with the instabilities on the liquid sheet amplifying, leading to the catastrophic breakdown of the liquid sheet into string-like ligaments and then further breakdown into droplets. The stable phase demonstrates the most consistent spray behaviour, although the breakup length remains erratic. In the collapsing stage, the swirling motion persists but with decaying energy. The instability amplitude on the liquid sheet reduces until it becomes a smooth tulip like shape, which eventually collapses. 

The transition between each stage is unclear as there is no immediate change in spray behaviour between events. While the change from the developing to the developed and stable stage is abrupt, the change from the latter to the collapsing stage is gradual. For consistency in determining the transition between each stage of all spray bottle actuations, four event markers were identified:
\begin{enumerate}[label=\roman*.)]
    \item \textit{Initial jet: }The first frame of developing flow was considered when  the first drop was visible from the nozzle tip. 
    \item \textit{Start of the stable stage}: This frame was identified by the formation of ligaments at its lowest point.
    \item \textit{Start of collapsing stage}: This was found by examining images in reverse at 25 frames per second from a known collapsing flow, then adjusting the start frame until no change was found in the spray profile. 
    \item \textit{Final droplet formation}: This was the last frame of the collapsing flow. As the cone collapses into a tubular jet, it is the first frame where the last formed droplet separates from the jet. This typically leaves behind a droplet that rests on the nozzle and one remains airborne. As the outlet orifice of these sprays is recessed within the nozzle, separation may occur within the nozzle tip but was only considered where visible.
\end{enumerate}

\begin{figure}[h]
\centering\includegraphics[width=1\linewidth]{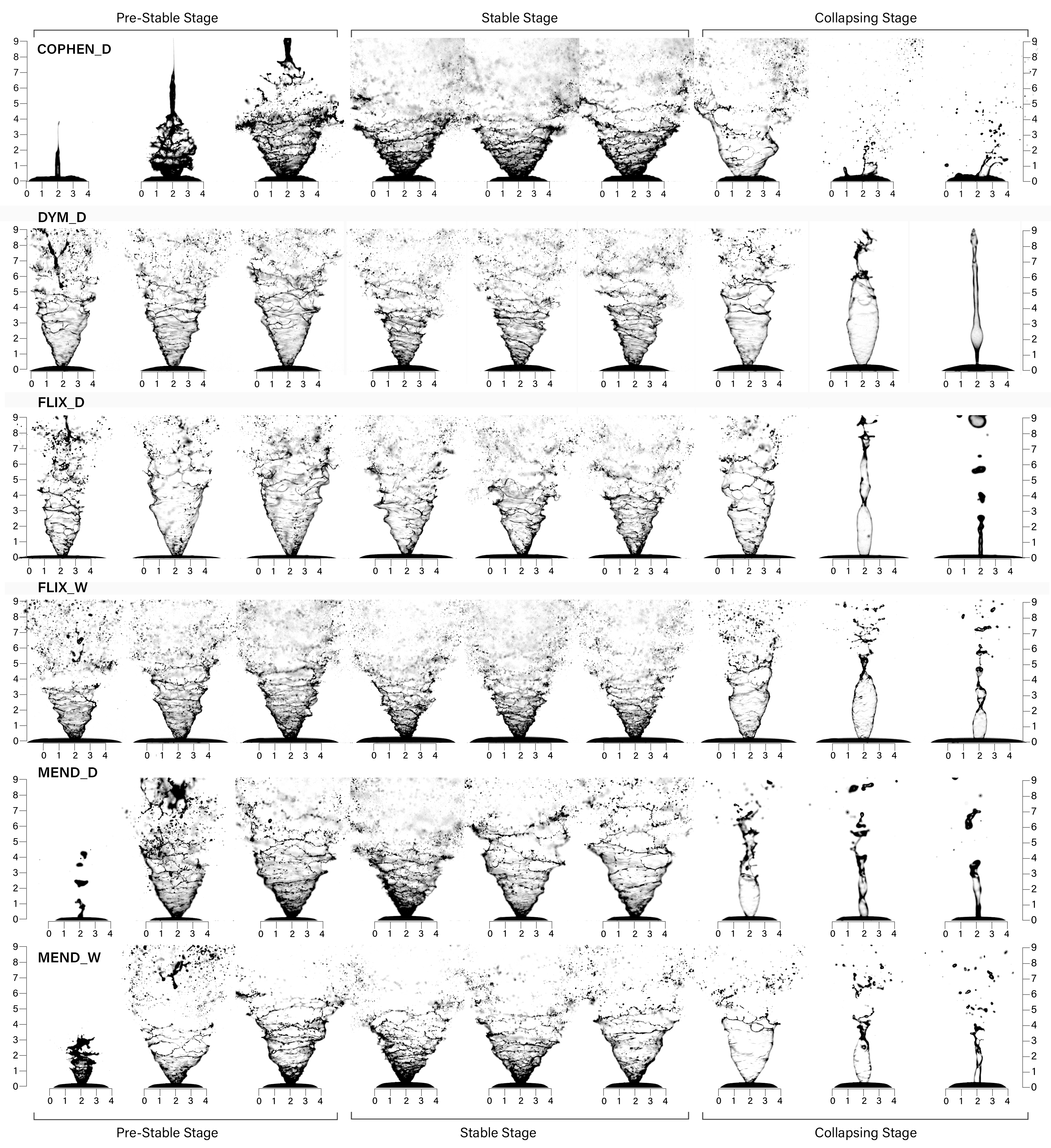}
\caption{A qualitative comparison of phases for all six nozzle and fluid combinations at the average adult actuation force (63 N). Each sample was chosen as the closest option to the mean stage length. The images were selected as 0.25, 0.50 and 0.75 times fraction of the stage duration. The axes units are in mm. }\label{fig:phaseComparison}
\end{figure}

From the preceding defined spray events, the pre-stable, stable, and collapsing stages were determined for all spray actuations. Figure \ref{fig:phaseComparison} presents three sample images (of each stage) taken at 0.25, 0.50 and 0.75 fraction of each stage length. The pre-stable events in all atomizers demonstrated vastly different behaviour. The COPHEN and MEND bottles exhibited wider sprays compared to the DYM and FLIX bottles. The drug formulation for the FLIX and MEND bottles formed longer breakup lengths compared to water (labelled with suffix '\_W'). Each stable stage image of each nozzle and fluid combination were similar except from the MEND bottle where the ligament formation was less prominent over time. During the collapsing stage, the atomization behaviour was similar for all nozzle and fluid combinations with the exception of COPHEN and this may be attributed to the nozzle which atomizing the fluid horizontally and hence the perpendicular gravity force on the collapsing cone caused it to disintegrate earlier compared to the other devices. 

\begin{figure}[h]
\centering\includegraphics[width=0.95\textwidth]{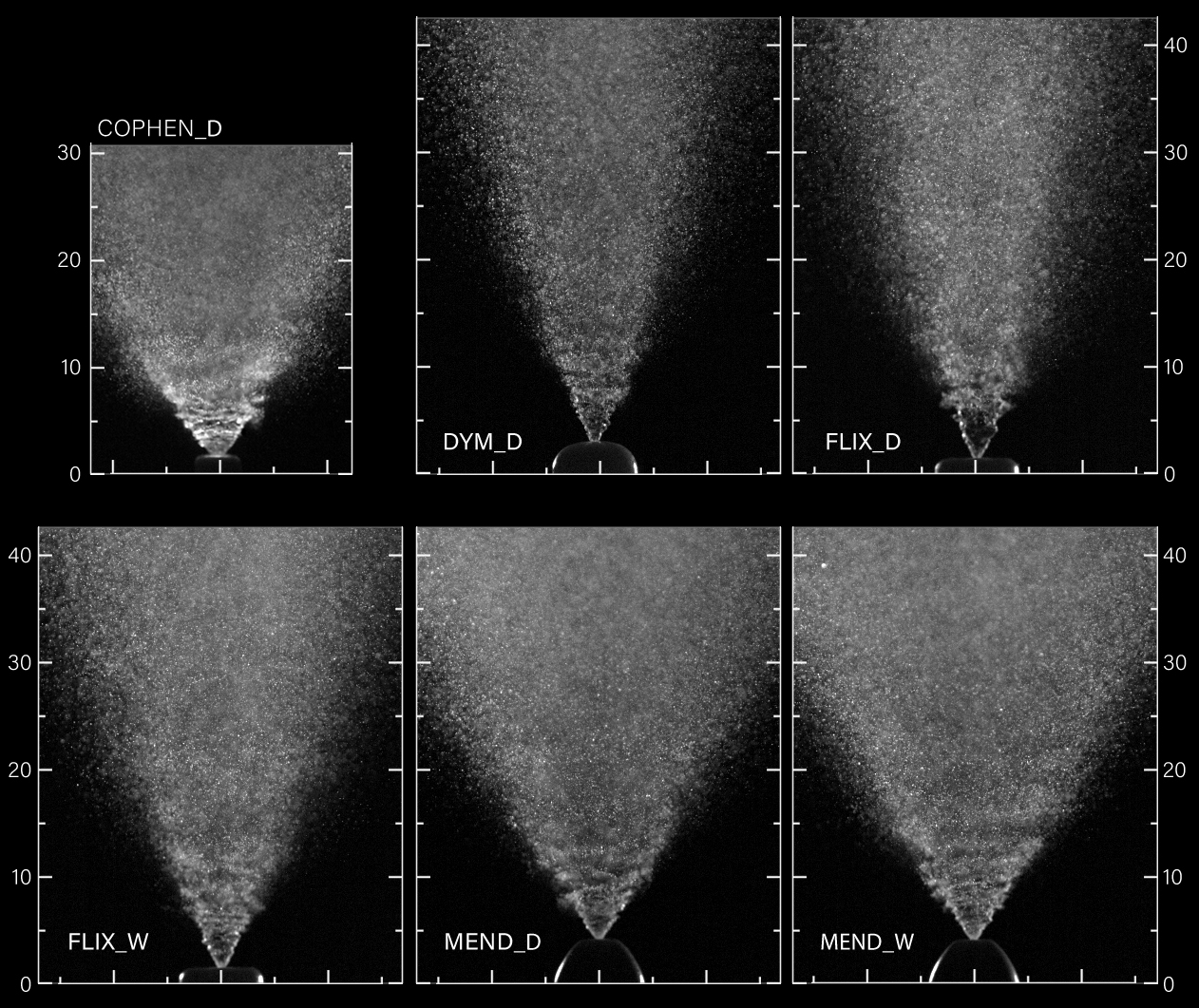}
\caption{Dark field images for a wider view of the atomized spray during the stable phase for each spray bottle. The dimensions of each image are in units of mm, where the images are limited to a height of 42mm and width of 33mm. The COPHEN\_D image is smaller due to the resolution captured for the spray.}
\label{fig:darkField}
\end{figure}

The dark field images provide a wider view perspective of the spray atomization in the ambient air during the stable stage. One image from each bottle is shown in Figure \ref{fig:darkField} which provides context of the spray breakup length which is relatively short compared to the entire spray plume. The spray plume width continued expanding and in some cases wider than the image frame of 33mm. The images show COPHEN and MEND bottles producing the widest spray plumes, while the DYM and FLIX bottles had the thinnest.

\begin{figure}[h]
	\centering\includegraphics[width=0.8\linewidth]{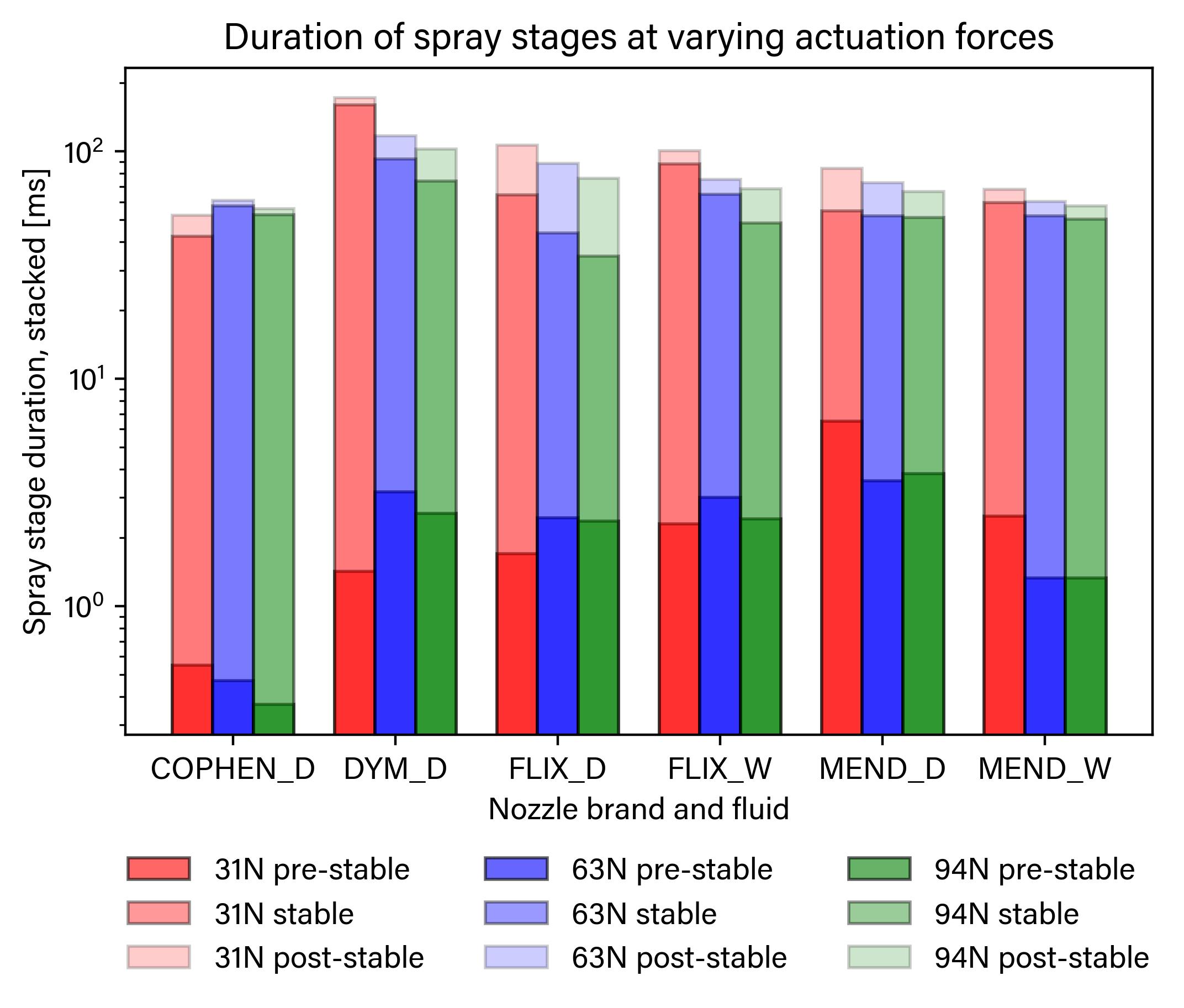}
	\caption{Cumulative (stacked) bar graph displaying the total duration of the spray atomization, contributed by each stage duration.}\label{fig:phaseDuration}
\end{figure}

Figure \ref{fig:phaseDuration} combines each stage duration to produce a cumulative bar graph of the total time taken for the atomization event. The DYM nozzle actuated at 31 N took the longest time to complete the atomization at 172~ms, while the shortest spray event was COPHEN actuated at 31 Newtons 52 ms, but this appears to be an outlier. For example, in all cases (except COPHEN), increasing the actuation force, decreased the spray duration thereby producing a rapid spray, as expected. The pre-stable stage is the shortest stage, while the stable stage is the longest. However, for the FLIX device with drug formulation at 63 and 94 N, the collapsing stage lasts longer than the stable stage. 

\subsection{Visualisation of Breakup Length, Spray Penetration, Spray Cone Angles} 
\begin{figure}[h]
\centering\includegraphics[width=0.8\linewidth]{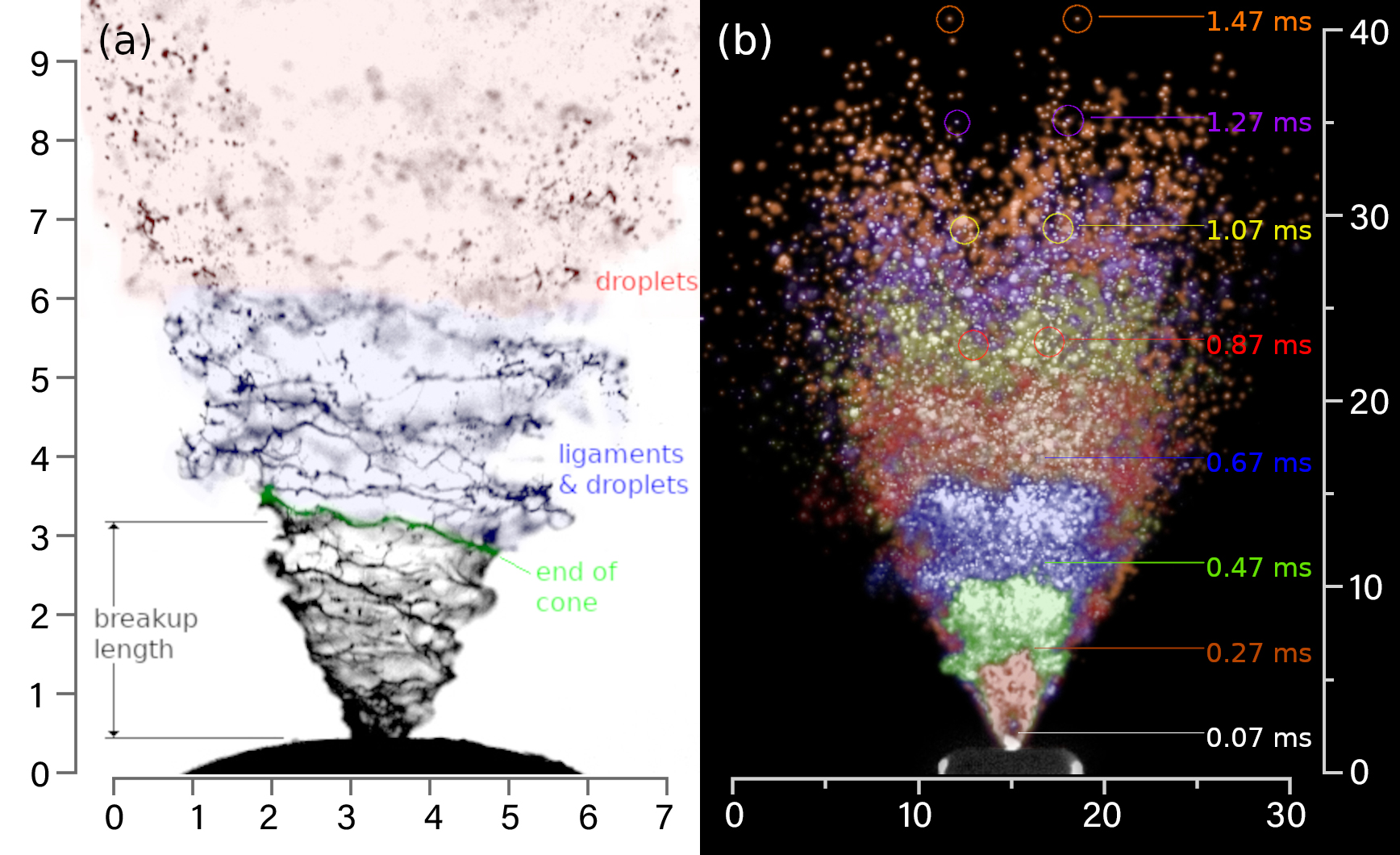}
\caption{(a) An example image using the Dymista bottle during the developed flow regime, actuated at 94 N, demonstrating the cone sheet, breakup length, ligaments and droplets. (b) An example image of the Flixonase bottle, actuated at 94 N, colour overlay of the initial spray droplets, where each tint represents a single frame of footage.}\label{fig:penOverlaybreakupLength}
\end{figure}

Figure \ref{fig:penOverlaybreakupLength} (a) demonstrates how the breakup length was interpreted, by highlighting the spray characteristics that is comprised of regions containing the cone sheet, ligaments, and droplets. The breakup length was determined by the image pixel height, noting that the green line highlights the end of the cone sheet (below the green line is the swirling cone sheet). As the liquid sheet is swirling, the green line highlighting the edge of the cone sheet conforms to disturbance waves on the swirling sheet, which is seldom perpendicular to the direction of flow, nor is it a straight line. Therefore, the breakup length measurement was defined as an average height based on a line of best fit. 

Figure \ref{fig:penOverlaybreakupLength} (b) provides an example of an image overlay of the spray penetration labelled with time taken to reach each location, taken from the Flixonase nozzle, actuated at 94 N. As each spray moves forward with time, the spray penetration and velocity can be calculated. The top four overlays include circles to indicate two droplets which had travelled furthest from the bulk spray.

\begin{figure}[h]
\centering\includegraphics[width=1\linewidth]{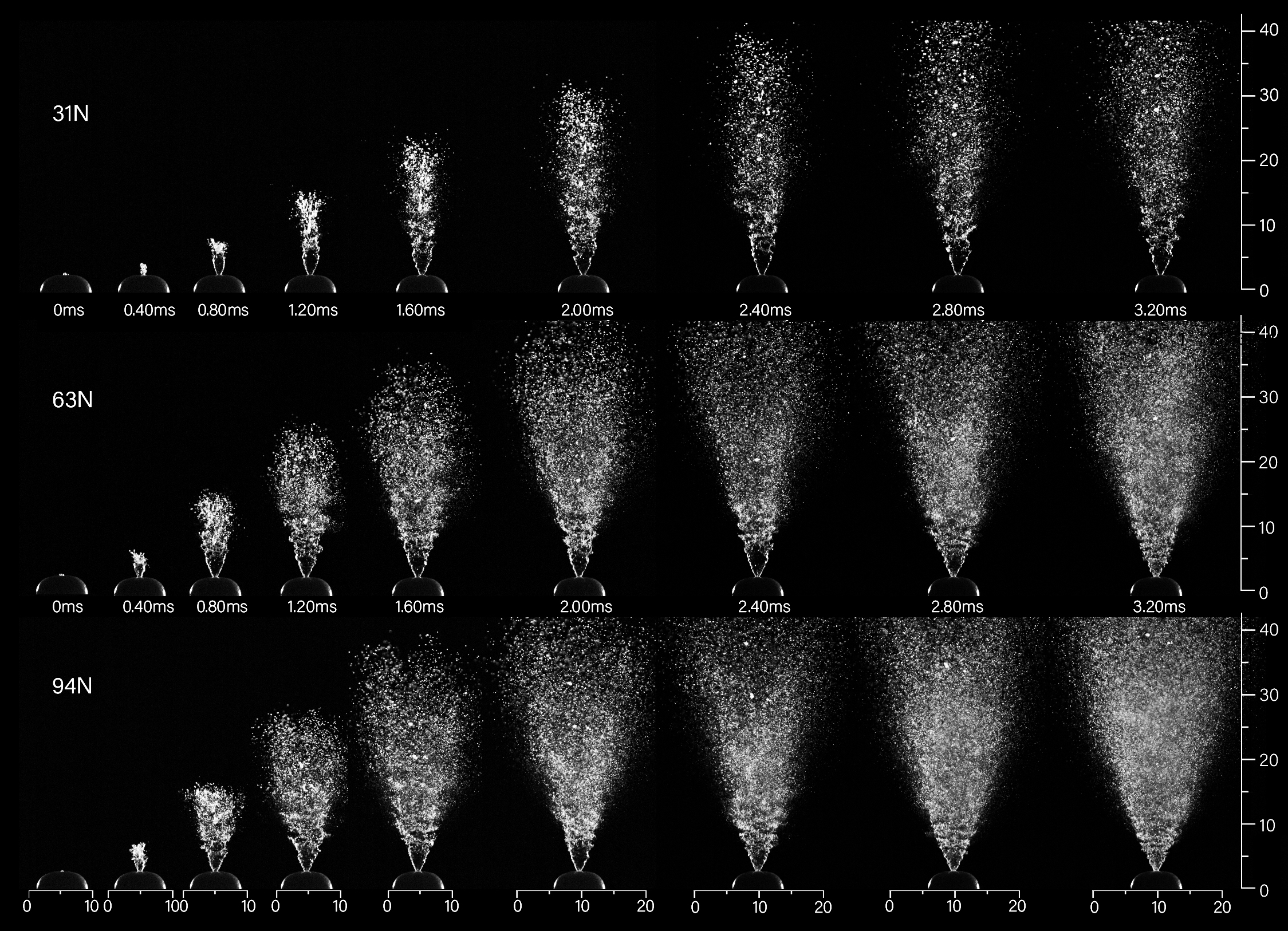}
\caption{Example images of the spray development during the pre-stable stage from the dark field images of the Dymista bottle (DYM) demonstrating the effect of the actuation force on spray penetration. The dimensions are in units of mm, where the images are limited to a height of 42mm and width of 33mm.}
\label{fig:darkFieldCompare}
\end{figure}

The spray development leading to its penetration is depicted in Figure \ref{fig:darkFieldCompare} for the DYM bottle at different actuation force. A low actuation force produced a weak atomization leading to a longer breakup length, thinner spray plume, and slower spray penetration. In contrast, the highest actuation force provided greater energy for a stronger atomization leading to a shorter breakup length, wider spray plume, and more rapid spray penetration. 

\begin{figure}[h]
\centering\includegraphics[width=0.8\linewidth]{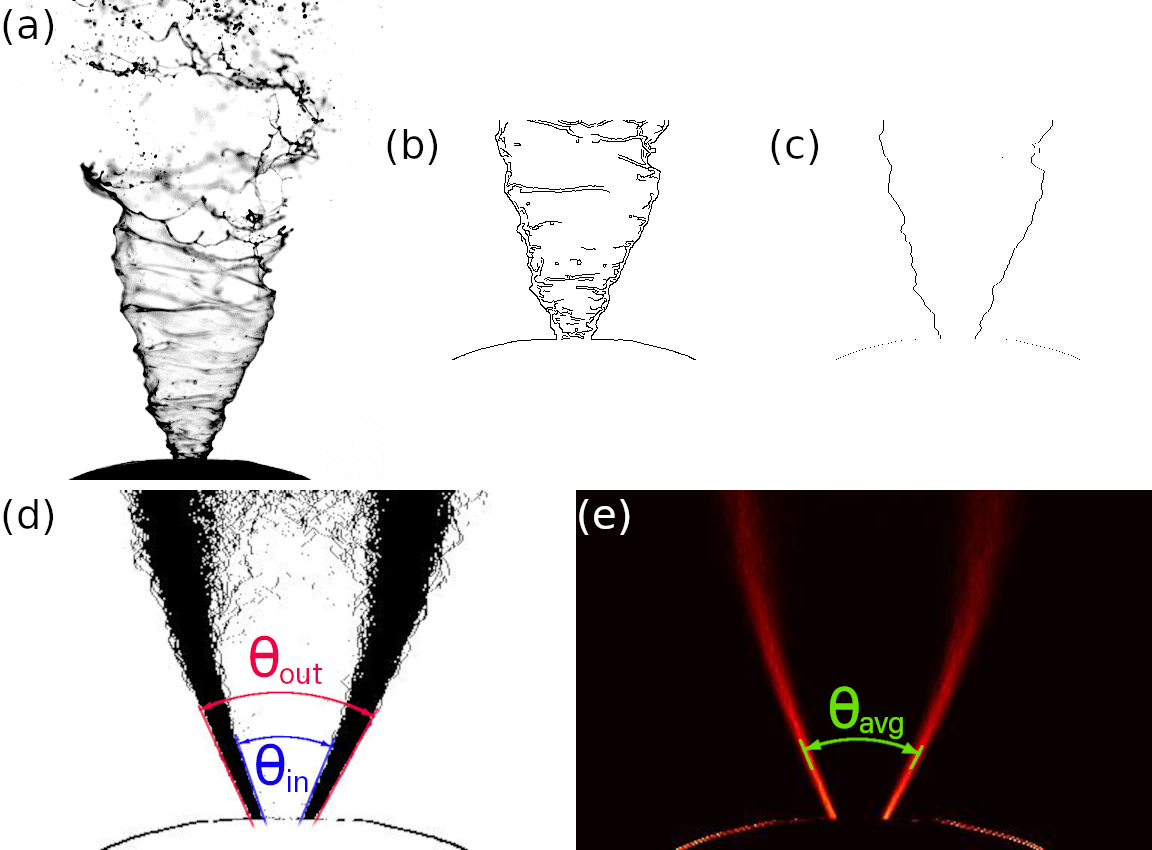}
\caption{Image processing steps to determine the spray plume edge. a) Original image; b) Image cropped Canny edge detection; c) Edge detection by taking the first and last pixel in each row; (d) Collection of all edge detected images processed from panel (c) and overlaid onto each other; e) Overlayed images as a heatmap to provide an average cone angle. Half cone angle = $\theta_{in}/2$, dispersion angle = $(\theta_{out} - \theta_{in})/2$ and $\theta_{avg}$ half cone angle = $\theta_{avg}/2$. }
\label{fig:cannyProcess}
\end{figure}

Figure \ref{fig:cannyProcess} provides the outputs from the image processing step to determine the spray plume edge. This figure also includes an example of where the measurements were taken for determining the half cone angle, where $\theta_c = 0.5\theta_{in}$ and the dispersion angle, $\theta_d = 0.5(\theta_{out} - \theta_{in})$, where $\theta_{out}$ and $theta_{in}$ are the outer and inner edges of the overlayed cone edge respectively (see \ref{fig:cannyProcess}). $\theta_{avg}$ is the average cone angle, found by overlaying the processed edge detected images onto a heatmap. 

\subsection{Statistical Analysis of Spray Characteristics} 
\begin{figure}[h]
	\begin{subfigure}[p]{0.475\textwidth}
		\centering
		\includegraphics[width=\textwidth]{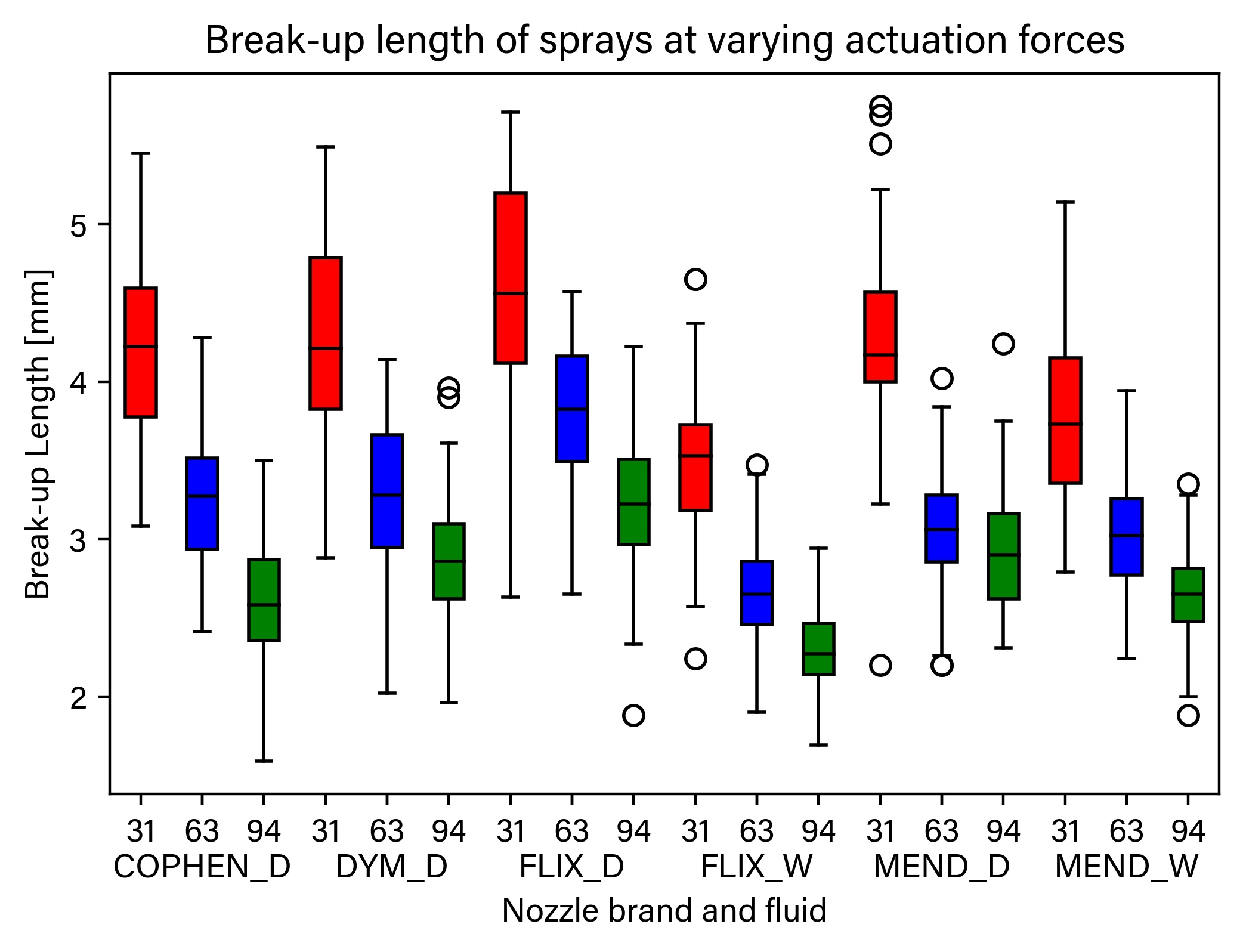}
		\subcaption{}
		\label{fig:bplotBUlength_a}
	\end{subfigure}
	\begin{subfigure}[p]{0.490\textwidth}
		\centering
		\includegraphics[width=\textwidth]{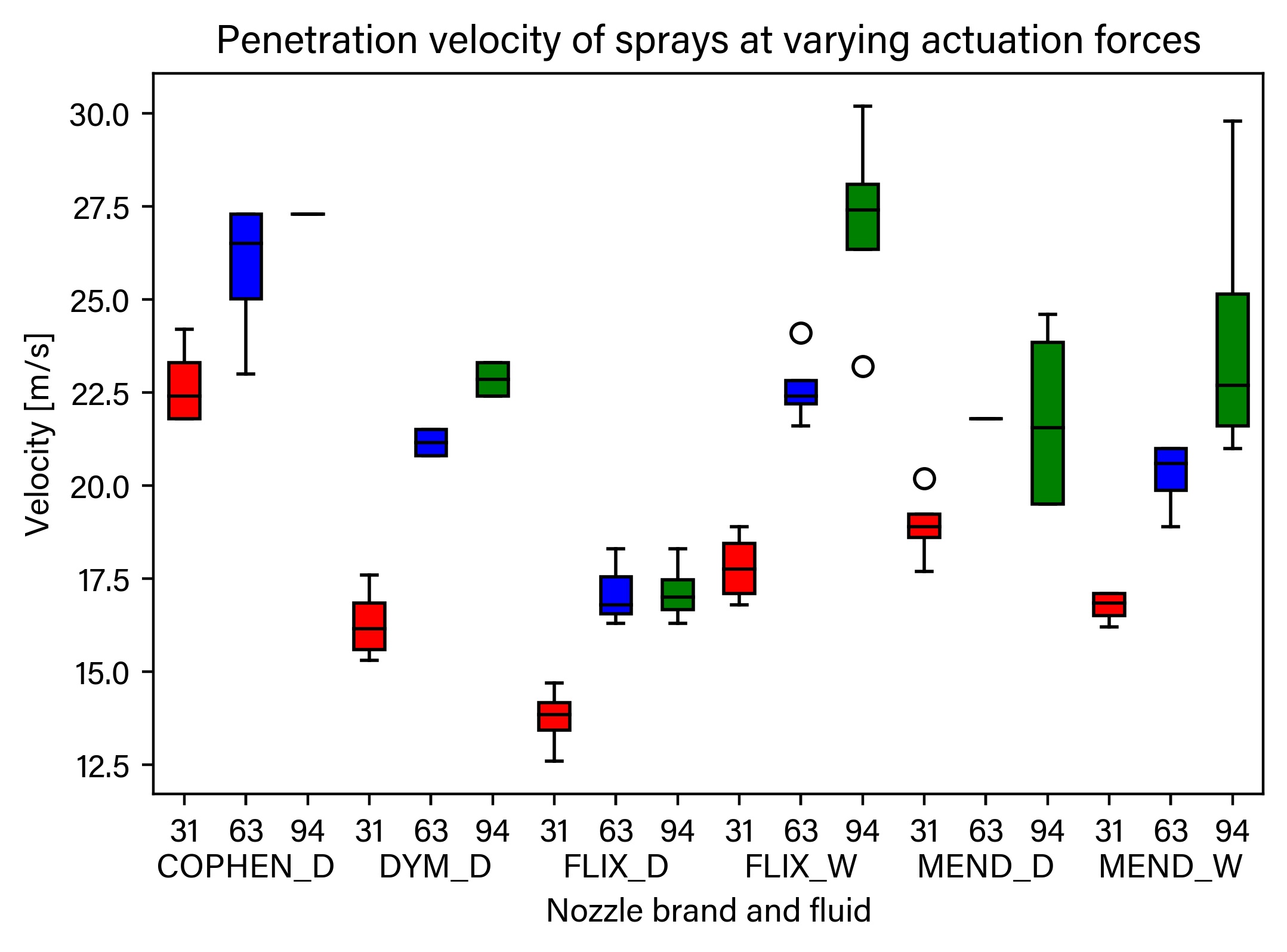}
		\subcaption{}
		\label{fig:bplotPenVel_b}
	\end{subfigure}
	\begin{subfigure}[p]{0.475\textwidth}
		\centering
		\includegraphics[width=\textwidth]{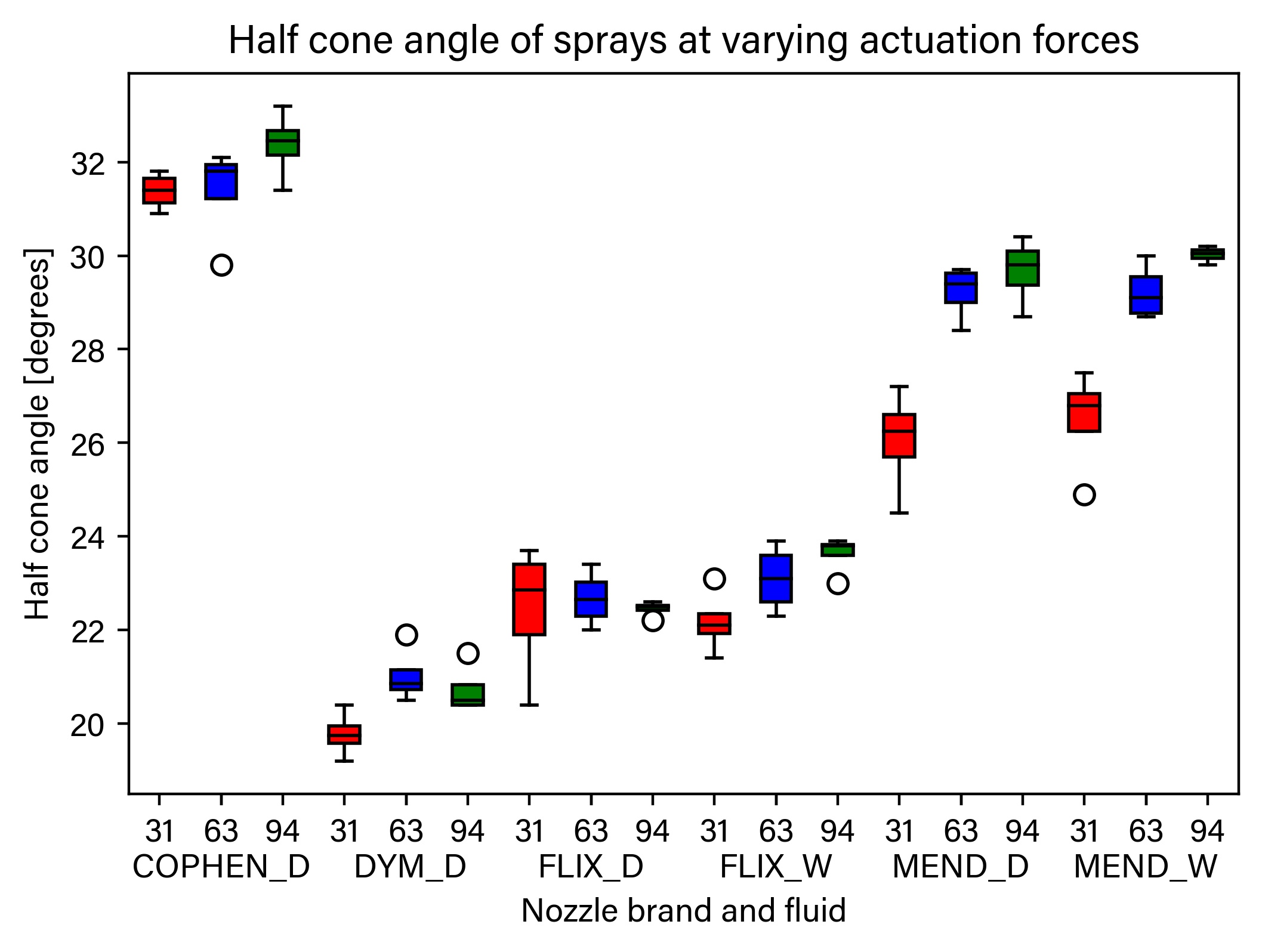}
		\subcaption{}
		\label{fig:bplotConeA_c}
	\end{subfigure}
	\begin{subfigure}[p]{0.475\textwidth}
		\centering
		\includegraphics[width=\textwidth]{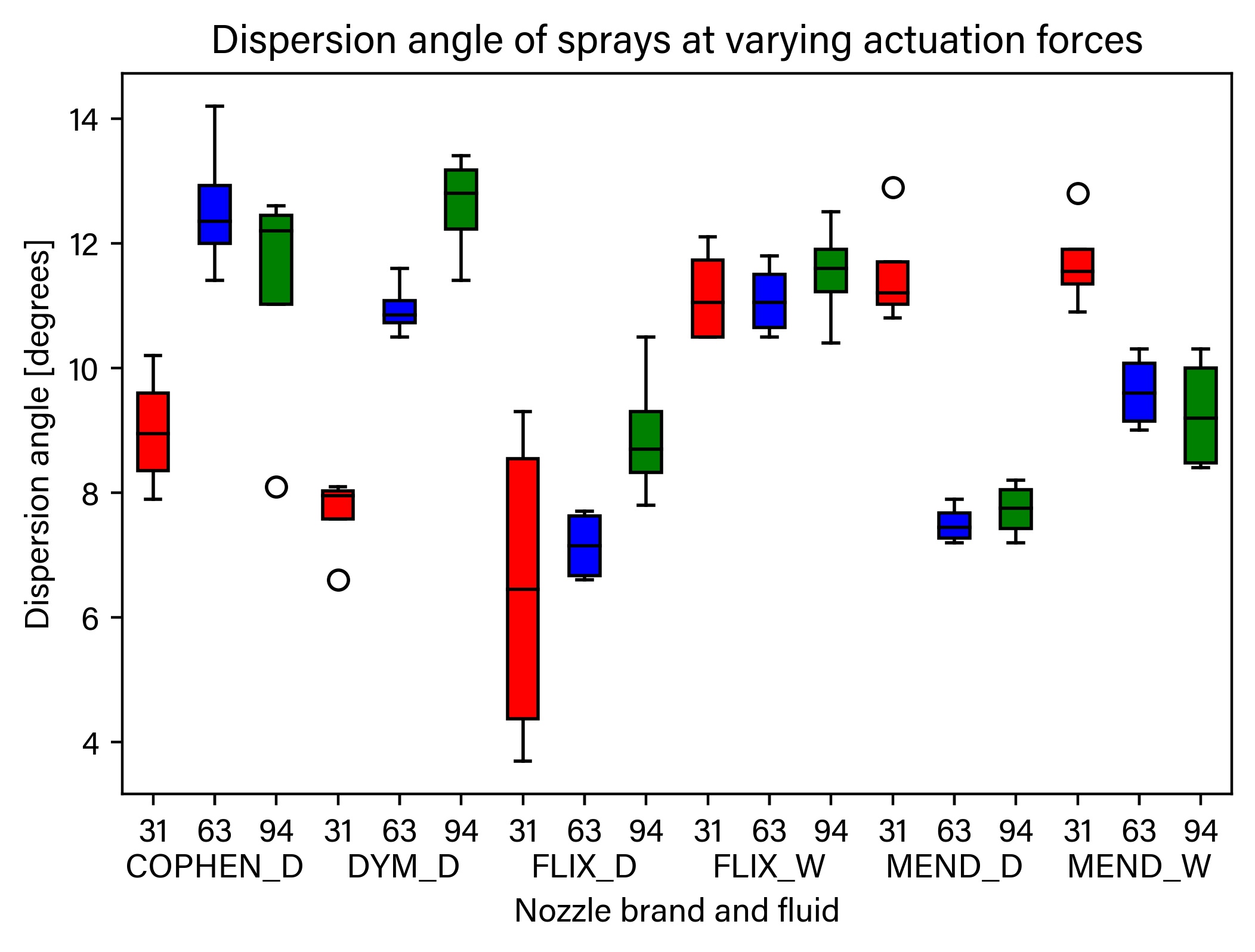}
		\subcaption{}
		\label{fig:bplotDispersion_d}
	\end{subfigure}
	\caption{Statistical analysis of spray atomization through box-plots to determine the breakup length, penetration velocity, half-cone angle, and dispersion angle. Circles denote outlier data, while some invariant data is shown with a horizontal line.}
	\label{fig:bplot}
\end{figure}

Tabulated data of all results are given in Table \ref{tab:Results1} and mean values in Table \ref{tab:Results2}. The statistical analysis of the data is given as boxplots.
Statistical analysis of the spray development was performed through box-plots given in Figure \ref{fig:bplot}. The breakup lengths ranged from  2.3 to 4.6~mm, with a mean of 3.3~mm. There was a clear relationship of decreasing breakup length with increasing actuation force (Figure \ref{fig:bplotBUlength_a}). All drug formulations produced similar results, with a maximum variation of 0.7~mm. The water events were significantly shorter than those using the drugs, implying a relationship to viscosity. 

The spray half-cone angle ranged from $19.8^\circ$ (DYM at 31 N) to $32.4^\circ$ (COPHEN at 94 N) with a mean value of 26.1$^\circ$ for drug formulation at 63 N. The narrowest plume found for the DYM\_D bottle, and the widest plumes from the COPHEN\_D device. In some instances, there is an increase in half-cone angles with increased actuation force, in particular there is most significant difference between 31 N and 64 N force for some of the sprays, however not in all cases.

The dispersion angle ranged from $6.5^\circ$ (FLIX\_D at 31 N) to $12.6^\circ$ (COPHEN at 31 N) with a mean of $9.5^\circ$ for drug formulation at 63 N. The dispersion angle demonstrated greater variation among the different spray devices. There was a general increase in dispersion angle for increased actuation force for all bottles except for the MEND bottle which exhibited the opposite relationship of decreased dispersion angle with increased actuation force.

The spray penetration velocity ranged from 13.8 m/s (FLIX\_D at 30 N) to 27.3 m/s (COPHEN at 94 N) with a mean of 21.7 m/s for drug formulation at 63 N. In general, there is an increase in spray penetration velocity with increased actuation force. Greater variation in velocity was found for the higher actuation force. 

\subsection{Droplet Size Distributions}
\begin{figure}[h]
\centering\includegraphics[width=0.9\linewidth]{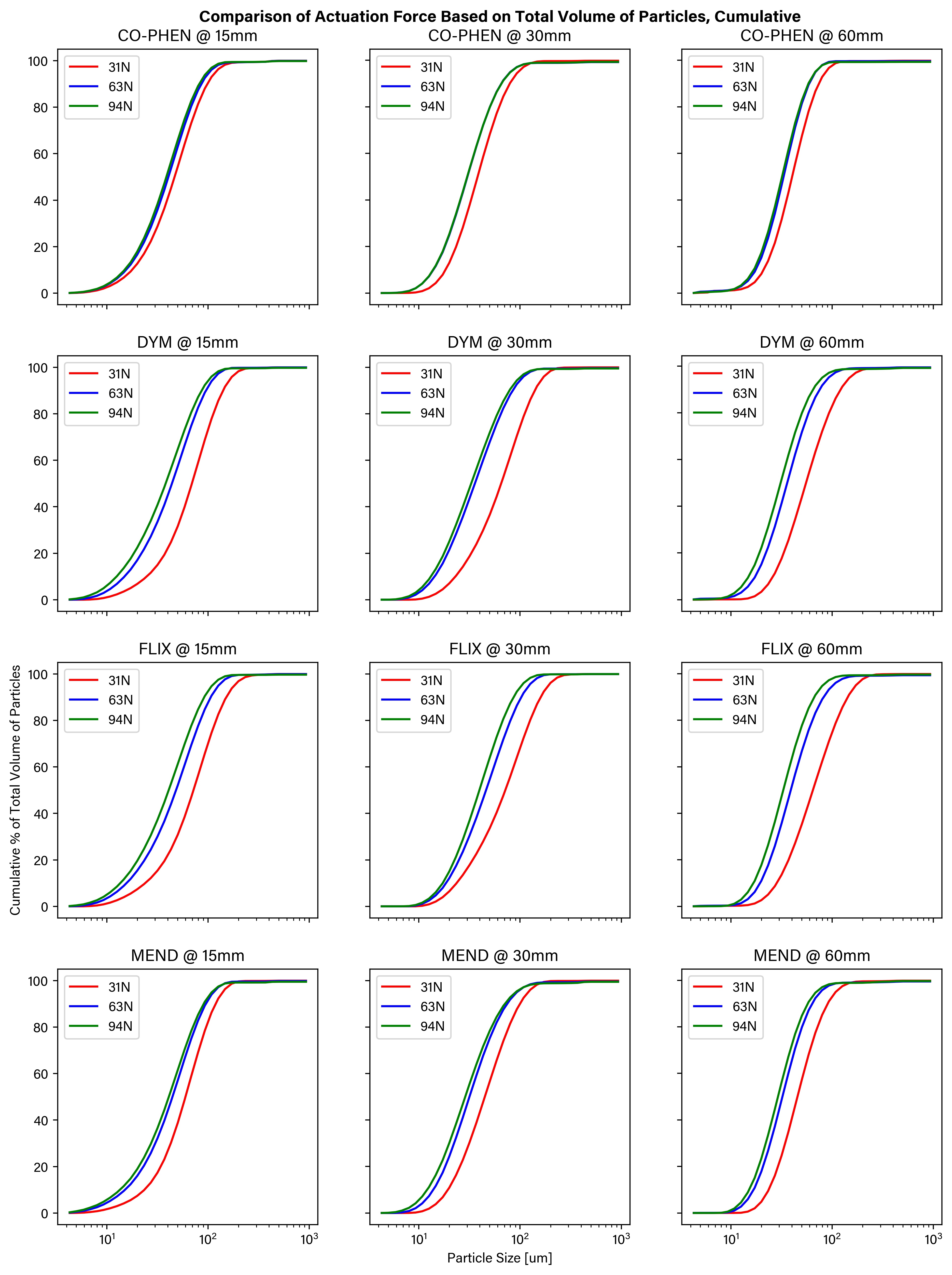}
\caption{Droplet size distribution comparing actuation forces.}
\label{fig:PressDSD}
\end{figure}

\begin{figure}[h]
\centering\includegraphics[width=0.9\linewidth]{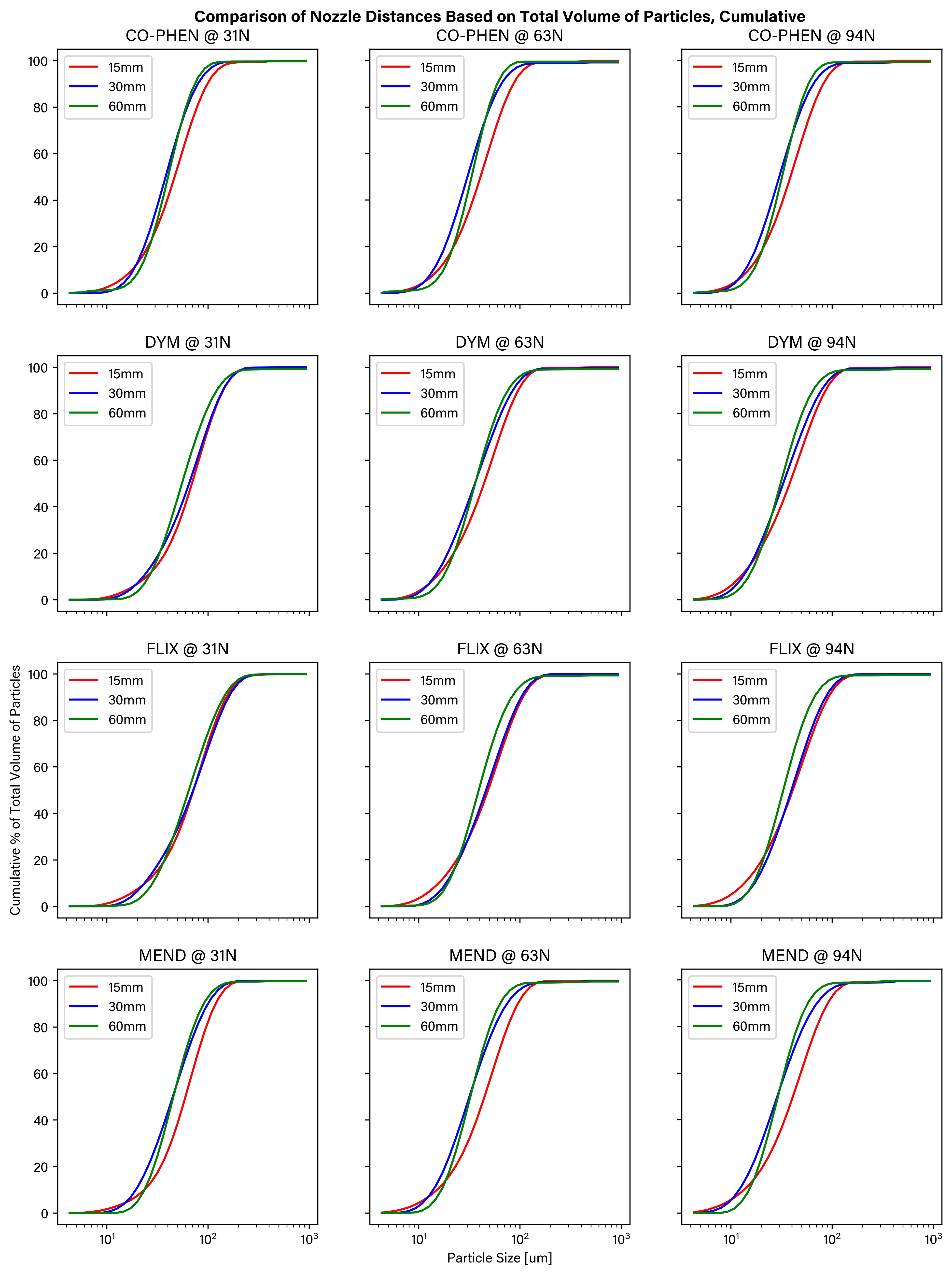}
\caption{Droplet size distribution comparing samples at varying distances to the nozzle tip.}\label{fig:distDSD}
\end{figure}

Droplet size distributions for the drug formulations are given in Figures \ref{fig:PressDSD} and \ref{fig:distDSD}. Each line represents the cumulative percent of the total volume of each particle size for a particular scenario. A steeper gradient implies a narrower size range (lower polydispersivity). Offsets to the left or right imply a greater number of smaller or larger particles respectively. Figure \ref{fig:PressDSD} compares the affect of actuation force on the size distribution. There is a general decrease in droplet diameters with increased actuation force, although there is negligible change for COPHEN at 63 and 94 N force suggesting it has reached its maximum application force. The decrease in droplet diameter size is most significant from 31 N to 63 N, while the decrease between 63 N to 93 N is more subtle.

Figure \ref{fig:distDSD} compares the droplet size distribution at varying distances from the tip of the nozzle which demonstrates the change in droplet diameters due to coalescence, evaporation and secondary breakup that may occur. In all samples taken at $60mm$, a sharper gradient was present compared to the 15 and 30~mm samples; meaning the range of droplets sizes is less diverse at a greater distance. In most of the cases, the droplet diameters decrease with measurements further downstream, although the variations are not always consistent among all bottles.

\begin{figure}[h]
\centering\includegraphics[width=0.9\linewidth]{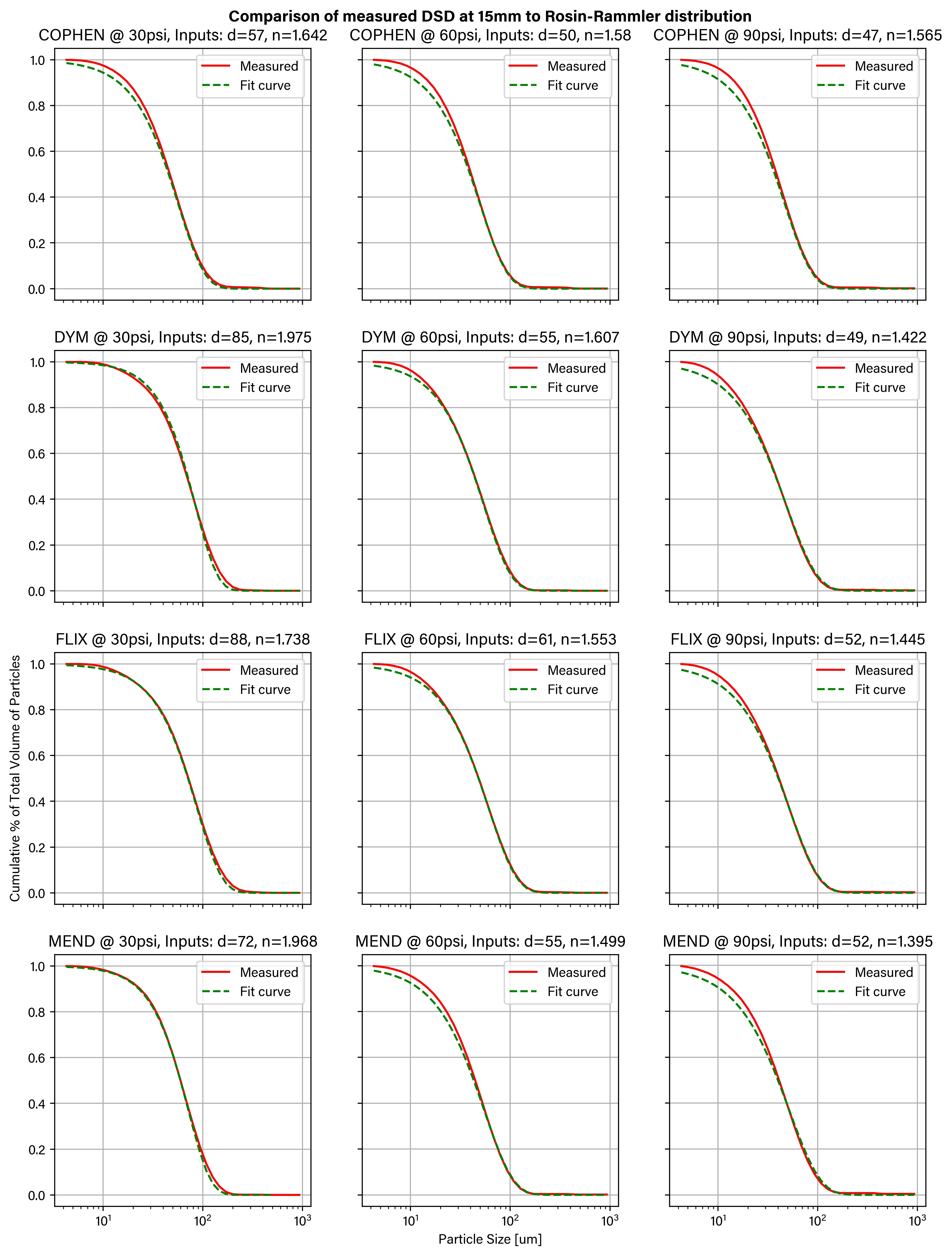}
\caption{Droplet size distribution comparing Rosin-Rammler distribution inputs to measured DSD at 15mm.}\label{fig:rrDSD}
\end{figure}

Figure \ref{fig:rrDSD} compares the measured droplet size distribution with recommended input values for a Rosin-Rammler distribution function which is a commonly used function in computational fluid dynamics modelling to represent a size distribution. The distribution function is defined as 
\[
Y_d = e^{-(d/\overline{d})^n}
\]
where $Y_d$ is the mass fraction of droplet of diameter greater than $d$. The inputs required are  $\overline{d}$ which is the size constant and $n$ which is the spread parameter. The measured distance of 15~mm from the nozzle tip was selected as it can be used as a suitable approximation for the droplet size distribution after primary breakup, as the initial droplet diameter conditions for CFD simulations.

\section{Discussion}
An allocated fluid volume is dispensed through the atomizer during a nasal spray pump actuation when fully primed. As the dispensed volume is predefined, an increase in actuation force decreases the completion time of a spray event (Figure \ref{fig:phaseDuration}). This trend was evident in all spray cases except for the Co-Phenylcaine bottle at 31 N actuation force. Similarly, \cite{Fung2013}, reported decreased spray duration for increased actuation force when comparing three different actuation forces on a simplex nozzle. \cite{Guo2006influence} investigated an electronic spray actuator and found that too low of an actuation force resulted in an incomplete stroke of the nasal spray. While the present study did not investigate the minimum actuation force required for atomization to occur, it stands to reason that an actuation force of 31 N was inadequate to overcome the resistance of the Co-Phenylcaine spray pump. 

An increase in actuation force led to a decrease in breakup length (Figure \ref{fig:bplotBUlength_a}) as the higher energy exerted by the actuation force translated to greater instability amplitudes. The breakup lengths ranged from $2.3$ to $4.6~mm$, which are within the range of existing data in the literature. \cite{shrestha2020primary} compared drug and water formulations in one bottle and found breakup lengths ranging from $3.4$ to $6.0~mm$, with a mean of $5.2~mm$. \cite{inthavong2012external} found a mean breakup length of $5.7~mm$. However, their definition was based on visual inspection of droplet formation and not at the sheet edge used in this study, which may contribute to the slightly higher figures. 

Several studies in the literature provide half cone angles \cite{Foo2007, Cheng2001, Suman2002,Fung2013, shrestha2020primary, kundoor2011effect}, ranging from $13^\circ$ to $37.5^\circ$ with a mean of $25.1^\circ$. In the present study, the half cone angles ranged from $19.8$ (DYM at 31 N) to $32.4^\circ$ (COPHEN at 94 N) with a mean of $25.8^\circ$. There was no clear relationship between actuation force and cone angle among the different devices. DYM\_D, FLIX\_W, MEND\_D and MEND\_W showed increased cone angle with increased actuation force. However, FLIX\_D demonstrated the opposite trend, while COPHEN\_D, FLIX\_D provided inconclusive results. 

There was no appreciable difference between water and drug formulation, indicating viscosity had negligible impact on cone angle. The lack of relationship between cone angles to fluid pressure and viscosity is consistent with inviscid theory of a single particle flowing through a simplex atomizer \cite{giffen1953atomization, rizk1985internal}, which indicates that cone angle is solely a function of nozzle geometry i.e., the ratio of the tangential and axial velocity components of the fluid as it leaves the outlet orifice. 

The dispersion angles varied from $6.5^\circ$ to $12.6^\circ$ with a mean of $9.9^\circ$ (Figure \ref{fig:bplotDispersion_d}). Previous work of \cite{shrestha2020primary} found a mean dispersion angle of $8.65^\circ$; however, their study was limited to a single bottle with variations only in liquid volume and formulation (drug or water). Currently, there is no other reported data for the dispersion angle from nasal sprays. Yet, this phenomenon is a required input for computational atomization primary breakup models such as the Linear Instability Sheet atomization (LISA) model. 

Dispersion angles arise from Kelvin-Helmholtz instabilities on the swirling liquid sheet. As the swirling liquid sheet moves through the ambient air, instability waves form due to the shear layers. It is expected that a greater actuation force with a higher energy input should create more aggressive instabilities and perturbations on the liquid sheet and hence, a greater dispersion angle. However, this correlation was only found for the Dymista and  Flixonase nozzles delivering the drug formulation. The lack of correlation in the other bottles appears to be a function of the breakup length. A decreased breakup length leads to a shorter cone sheet and causes ligaments to form sooner. The hollow swirling cone sheet formed larger waves (in amplitude and wavelength) at lower pressures than the shorter cone sheet found for higher pressures. This, in turn, creates a more well-defined and consistent cone edge, even with the prominent ligament formation. The variation in dispersion angle in the Flixonase bottle with drug formulation compared to water is noteworthy, as it reflects the idea that an increase in viscosity causes a decrease in the development of Kelvin-Helmholtz instability perturbations.

The $\theta_{avg}$ half cone angles defines the spray plume and gives the value for the spray cone angles in CFD. All $\theta_{avg}$ half cone angles lie within the range of their respective minimum and maximum half cone angles, to derive the dispersion angles resulting from perturbations of the liquid sheet oscillating about $\theta_{avg}$.

The variations in droplet size distribution (Figure \ref{fig:PressDSD}) demonstrated that, in general, a lower actuation force yielded a larger droplet size which corroborates with the atomization concept; that kinetic energy is used to transform bulk fluid into smaller droplets. In all cases, the 31 N case representing an average child's actuation force \cite{Doughty2011} yielded larger droplets compared to higher actuation forces. Co-phenylcaine is unique in its droplet size distribution which remained relatively unchanged between 63 and 94N actuation force, where these forces represent the average and maximum adult actuation force \cite{Doughty2011}. It also has the most consistent droplet size distribution across the actuation forces tested. The device was from a Co-phenylcaine Forte spray where it has multiple uses for anaesthetic and vasoconstrictor for nasal tissue. It was connected with a Flexi Nozzle (ENT Flo\textsuperscript{\textregistered}) long tube inserted deep into the nasal cavity, and the atomized spray is produced parallel to the tube (perpendicular to the bottle).  

The droplet size distribution variation as the droplets travel downstream of the nozzle showed the distribution spread narrows with downstream position, evident by the DSD curve at 60mm exhibiting the steepest gradient (Figure \ref{fig:distDSD}). The narrowing distribution spread is most likely due to secondary breakup of the droplets, where the relative air velocity deforms larger droplets into smaller ones \citep{guildenbecher2009secondary}. The DSD at 60mm for all bottles except the Flixonase at 94 N actuation showed a reduced volume of droplets in the range of $10$ to $30~\upmu$m diameter, as shown by the lower portion of the 60mm curve (green colour) having a longer horizontal line until $30~\upmu$m diameter. There are two potential mechanisms to the reduction of smaller droplets: coalescence or evaporation. \cite{lefebvre2017atomization} suggests the likeliness of coalescence occurring is based on the number density of the spray and the time available for collisions. As the Malvern Spraytec measures DSD as a function of total volume, it is difficult to determine what phenomenon causes this result without knowing the total volume. Calculations for the time of complete evaporation of a single pure water droplet exist in the literature \cite{morawska2006droplet, redrow2011modeling}. Both models compare evaporation time for 1, 10 and $100~\upmu$m diameter droplets, each at 0, 20, 60 and 80\% relative humidity. Pure water droplets at $1~\upmu$m completely evaporate between 0.7 and 5.4 milliseconds, at $10 \upmu$m, evaporation time varies between 0.08 to 0.48 seconds and droplets at $100 \upmu$m can last between 7 and 45 seconds.
Dealing with spray velocities in the range of 15 to 30 m/s means the time for the droplets to travel from $15$ to $60~mm$ from the nozzle tip will be in the range of 1 to 3 milliseconds, making it unlikely that any droplets over $10 \upmu$m evaporate within the measured spray distances. However, this cannot be treated as conclusive because the initial droplet temperature of the models is greater than room temp. The Weber number ($We = \rho_A U_R^2 D / \sigma$) is a ratio of aerodynamic forces to surface tension force on a droplet, where $\rho_A$ is the density of air, $U_R$ is the relative velocity of the droplet and air, $D$ is the droplet diameter and $\sigma$ is the surface tension of the droplet. Using values available in the literature (\cite{han2016surface, eccleston2000rheological}), a Weber number of 3.5 was calculated for droplets at $150~\upmu$m and $27~m/s$. Various sources in \cite{lefebvre2017atomization} state that droplet breakup (atomization secondary breakup) will occur between Weber numbers of 10 to 22. With a low Weber number, it seems more plausible that partial evaporation of the larger droplets is the cause of the steeper gradient. While the rate of evaporation from \cite{morawska2006droplet, redrow2011modeling} indicate evaporation is not likely, the results from the literature were calculated under different velocity and temperature conditions.  

As a summary of the DSD produced from nasal sprays, Rosin-Rammler  functions were produced to match the distribution which serve as useful benchmark data for validating computational modelling of sprays.
\section{Conclusion}
Experimental investigations were performed on four commercial nasal sprays, with the original drug formulation and where possible, replaced with water. These experiments were performed at the mean paediatric, mean adult and maximum adult actuation forces and included high-speed videography and finding droplet size distribution through laser diffraction. The high-speed videography results were image processed to find phase durations, breakup lengths, spray cone angles and dispersion angles. It was concluded that actuation force affected phase duration, DSD and breakup length yet had no clear relationship to cone angle or dispersion angle. Viscosity appeared to affect dispersion angle and breakup length only; however, the relationship of viscosity on DSD was not investigated. Each nasal spray was subjected to evaporation, coalescence and secondary breakup, demonstrated by the DSD of each spray at varying distances to the nozzle tip, however there is contention with existing theory on the shift in DSDs. In addition to these conclusions, the study provides benchmark experimental data for future CFD investigation and inputs for atomization sub-models, in particular the dispersion angle which is rarely reported, and Rosin-Rammler functions for the DSDs.  

\section{Acknowledgements}
The authors would like to thank Deakin University for the PhD scholarship and start-up funds to support this project. 

\bibliography{main}

\newgeometry{a4paper, landscape, left=0.5in, right=0.5in, top=0.5in, bottom=0.5in}
\begin{landscape}
\begin{table}[h]
\resizebox{1.3\textwidth}{!}{
\centering
\begin{tabular}{ p{2.5cm} p{1.8cm} p{2cm} p{1.8cm} p{1.8cm} p{1.8cm} p{1.8cm} p{2cm} p{2.5cm} p{2.5cm} p{2.5cm} p{2cm} p{3cm} p{2cm} p{2cm} p{2.5cm} p{2.5cm}}
\hline
\textbf{Spray \newline device} & \textbf{Contents} & \textbf{Actuation Force} & \textbf{Half\newline cone \newline angle} & \textbf{Dis-\newline persion\newline angle} &  \textbf{$\theta_{avg}$ half cone\newline angle} &  \textbf{breakup length [mm]} & \textbf{Pre-stable duration [mS]} & \textbf{Stable \newline duration \newline [mS]} & \textbf{Collapsing duration [mS]} & \textbf{Penetration\newline speed\newline [m/s]} & \textbf{Size\newline Constant\newline [\=d]} & \textbf{Size \newline distribution \newline parameter [n]} & \textbf{Dv(10)\newline at 15mm \newline [$\mu m$]} & \textbf{Dv(50)\newline at 15mm \newline [$\mu m$]} & \textbf{Dv(90)\newline at 15mm \newline [$\mu m$]} & \textbf{SMD \newline at 15mm \newline [$\mu m$]}\\
\hline
Co-             & Drug & 31N & 31.4 (0.4) & 9.0 (1.0)  & 38.5 (0.5)  & 4.2 (0.6) & 0.55 (0.08) & 41.75 (4.60) & 10.15 (3.32) & 22.7 (1.2) & 57 & 1.642 & 19.5 (0.6) & 51.1 (1.9) & 107.2 (10.2) & 37.0 (2.5) \\
Phenylcaine     &      & 63N & 31.4 (1.1) & 12.6 (1.2) & 36.2 (1.0)  & 3.2 (0.4) & 0.47 (0.05) & 75.30 (0.76) & 3.08 (0.79) & 25.8 (2.0) & 50 & 1.512  & 16.6 (1.3) & 43.6 (3.0) & 91.0 (5.8)   & 32.6 (2.4) \\
(COPHEN\_D)     &      & 94N & 32.4 (0.8) & 11.3 (2.2) & 37.6 (1.4)  & 2.6 (0.4) & 0.37 (0.04) & 52.47 (0.64) & 3.33 (0.14) & 27.3 (0.0) & 47 & 1.565  & 16.8 (1.5) & 43.2 (3.7) & 90.0 (7.3)   & 32.0 (2.7)  \\
  \\
Dymista & Drug  & 31N & 19.8 (0.5) & 7.7 (0.7)  & 24.8 (0.5)  &  4.2 (0.6) & 1.42 (0.23) & 158.50 (6.38) & 12.03 (1.49) & 16.3 (1.0) & 85 & 1.975 & 26.3 (1.2) & 74.1 (3.9) & 149.7 (8.6)  & 52.7 (2.1) \\
(DYM\_D)     &  & 63N & 21.0 (0.6) & 11.0 (0.5) & 26.4 (2.6)  &  3.3 (0.4) & 3.18 (0.14) & 89.62 (1.99)  & 24.42 (1.66) & 21.1 (0.4) & 55 & 1.607 & 16.0 (0.3) & 47.0 (1.1) & 101.2 (2.8)  & 33.2 (0.9) \\
            &   & 94N & 20.7 (0.6) & 12.6 (0.9) & 28.3 (1.9)  &  2.9 (0.4) & 2.55 (0.21) & 71.53 (0.86)  & 28.17 (0.91) & 22.8 (0.5) & 49 & 1.422 & 13.2 (0.6) & 40.3 (1.5) & 90.7 (2.7)   & 27.1 (1.1) \\
  \\
Flixonase & Drug  & 31N & 22.4 (1.5) & 6.5 (2.8)  & 27.2 (2.1)  & 4.6 (0.7) & 1.70 (0.78) & 62.90 (7.80) & 42.23 (17.97) & 13.8 (0.9) & 88 & 1.738 & 25.4 (0.7) & 76.2 (2.1) & 158.0 (6.6)  & 52.5 (3.5) \\
(FLIX\_D)      &  & 63N & 22.7 (0.6) & 7.2 (0.6)  & 26.3 (0.5)  & 3.8 (0.4) & 2.45 (0.15) & 41.42 (1.03) & 44.97 (4.16)  & 17.1 (1.0) & 61 & 1.553 & 16.8 (0.6) & 51.5 (1.5) & 112.5 (3.1)  & 35.2 (1.2) \\
               &  & 94N & 22.5 (0.2) & 8.9 (1.2)  & 26.3 (1.4)  & 3.2 (0.4) & 2.37 (0.13) & 32.28 (0.70) & 41.38 (2.32)  & 17.2 (0.8) & 52 & 1.445 & 14.3 (0.5) & 42.8 (1.8) & 95.2 (4.2)   & 28.7 (1.6) \\
  \\ 
Flixonase & Water  & 31N & 22.2 (0.7) & 11.2 (0.8) & 27.6 (1.3)  & 3.5 (0.5) & 2.30 (0.56) & 85.65 (2.07) & 12.93 (1.31) & 17.8 (1.0) & - & - & - & - & - & -\\
(FLIX\_W)       &  & 63N & 23.1 (0.7) & 11.1 (0.6) & 28.7 (0.3)  & 2.7 (0.3) & 3.00 (0.30) & 61.77 (1.82) & 10.65 (1.64) & 22.6 (1.1)  & - & - & - & - & - & -\\
                &  & 94N & 23.6 (0.4) & 11.5 (0.9) & 28.8 (2.1)  & 2.3 (0.3) & 2.42 (0.18) & 46.17 (0.64) & 20.02 (0.73) & 27.1 (2.9) & - & - & - & - & - & -\\
  \\
 Mendeleev & Drug & 31N & 26.0 (1.1) & 11.5 (0.9)  & 32.7 (1.6)  & 4.2 (0.6) & 6.52 (0.98) & 48.30 (4.54) & 29.50 (6.47) & 18.9 (1.0) & 72 & 1.968 & 25.0 (1.1) & 63.8 (1.7) & 125.8 (4.1)     & 53.2 (9.4) \\
(MEND\_D)      &  & 63N & 29.3 (0.6) & 7.5 (0.3)   & 32.9 (0.6)  & 3.1 (0.4) & 3.55 (0.41) & 48.50 (0.72) & 20.83 (1.48) & 21.8 (0.0) & 55 & 1.499 & 22.7 (9.7) & 64.4 (15.1) & 132.1 (23.1)   & 27.9 (4.3) \\
               &  & 94N & 29.7 (0.7) & 7.7 (0.5)   & 34.0 (1.9)  & 2.9 (0.4) & 3.82 (0.60) & 47.47 (0.93) & 15.32 (1.41) & 21.8 (2.7) & 52 & 1.395 & 21.9 (10.4) & 63.9 (16.5) & 132.3 (24.9)  & 29.3 (1.7) \\
  \\
Mendeleev & Water & 31N & 26.5 (1.1) & 11.7 (0.8) & 32.4 (1.3)   & 3.8 (0.5) & 2.48 (0.85) & 57.13 (1.70) & 8.68 (1.56) & 16.8 (0.5) & - & - & - & - & - & - \\
(MEND\_W)      &  & 63N & 29.2 (0.6) & 9.6 (0.6)  & 33.8 (1.1)   & 3.0 (0.4) & 1.33 (0.56) & 50.77 (0.70) & 8.20 (0.61) & 20.2 (1.0) & - & - & - & - & - & - \\
               &  & 94N & 30.0 (0.2) & 9.3 (1.0)  & 34.6 (0.7)   & 2.7 (0.3) & 1.33 (0.66) & 48.97 (0.70) & 7.18 (0.22) & 24.0 (4.0) & - & - & - & - & - & - \\
\hline
\end{tabular}}
\caption{Spray cone angles, breakup lengths, phase durations, penetration speeds and the Rosin-Rammler parameters for each actuation force, spray device and spray content. All angles in degrees. Figures in parentheses are the standard deviations of the data set. Refer to figure \ref{fig:cannyProcess} for definition of cone angles.}\label{tab:Results1}
\end{table}

\newpage
\begin{table}[h]
\resizebox{1.3\textwidth}{!}{
\centering
\begin{tabular}{ p{2.5cm} p{1.8cm} p{1.8cm} p{1.8cm} p{1.8cm} p{1.8cm} p{2cm} p{2.5cm} p{2.5cm} p{2.5cm} p{2.5cm} p{2.5cm} p{2.5cm} p{2.5cm}}
\hline
\textbf{Actuation \newline Force} & \textbf{Contents} & \textbf{Half\newline cone \newline angle} & \textbf{Dis-\newline persion\newline angle} & \textbf{$\theta_{avg}$ half\newline cone \newline angle}&  \textbf{breakup length [mm]} & \textbf{Pre-stable duration [mS]} & \textbf{Stable \newline duration \newline [mS]} & \textbf{Collapsing duration [mS]} & \textbf{Penetration speed [m/s]} & \textbf{Dv(10)\newline at 15mm \newline [$\mu m$]} & \textbf{Dv(50)\newline at 15mm \newline [$\mu m$]} & \textbf{Dv(90)\newline at 15mm \newline [$\mu m$]} & \textbf{SMD \newline at 15mm \newline [$\mu m$]}\\
\hline
31N & Drug  & 24.9 (4.6) & 8.7 (2.4)  & 30.1 (4.5) & 4.3 (0.6) & 2.55 (2.47) & 77.86 (49.02) & 23.48 (16.17) & 17.9 (3.5) & 24.1 (0.9) & 66.3 (2.4) & 135.2 (7.4)  & 49.6 (6.0) \\
        
63N & Drug  & 26.1 (4.5) & 9.5 (2.5)  & 30.5 (4.5) & 3.4 (0.5) & 2.41 (1.25) & 59.21 (19.07) & 23.33 (15.51) & 21.7 (3.2) & 18.0 (3.0) & 51.6 (5.2) & 109.2 (8.7)  & 35.7 (7.8) \\

94N & Drug  & 26.3 (5.0) & 10.1 (2.3) & 31.5 (4.7) & 2.9 (0.5) & 2.28 (1.31) & 50.94 (14.50) & 22.05 (14.72) & 22.3 (3.9) & 16.5 (3.2) & 47.5 (5.9) & 102.1 (9.8)  & 32.0 (9.6) \\

31N & Water & 24.3 (2.4) & 11.4 (0.8) & 30.0 (2.7) & 3.7 (0.5) & 2.39 (0.67) & 71.39 (15.34) & 10.81 (2.63)  & 17.3 (0.9) & - & - & - & -  \\

63N & Water & 26.1 (3.3) & 10.4 (1.0) & 31.3 (2.8) & 2.8 (0.4) & 2.17 (0.98) & 56.27 (6.02)  &  9.43 (1.74)  & 21.4 (1.6) & - & - & - & -  \\

94N & Water & 26.8 (3.4) & 10.4 (1.5) & 31.7 (3.2) & 2.5 (0.4) & 1.88 (0.73) & 46.65 (2.55)  & 13.60 (6.88)  & 25.5 (3.6) & - & - & - & -  \\

\hline
\end{tabular}}
\caption{Mean spray cone angles, breakup lengths and phase durations of all nozzles at each actuation force and spray content. All angles in degrees. Figures in parentheses are the standard deviations of the data set. }\label{tab:Results2}
\end{table}

\end{landscape}
\restoregeometry

\newpage

\clearpage

\clearpage

\clearpage

\clearpage

\clearpage

\clearpage

\newpage

\clearpage

\clearpage

\newpage
\newpage

\begin{figure}[h]
\centering\includegraphics[width=0.9\linewidth]{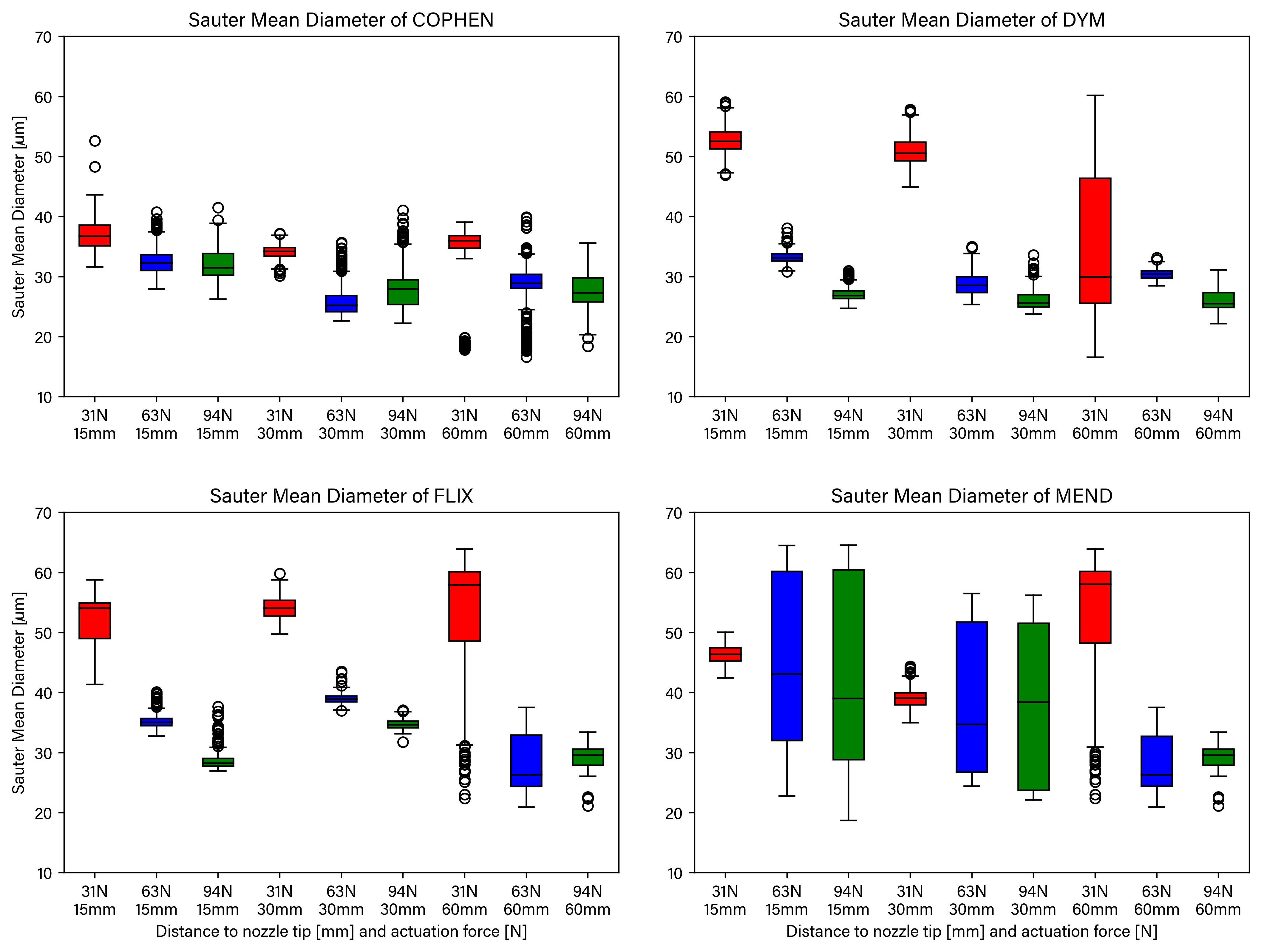}
\caption{Box plots for Sauter Mean Diameter of each nozzle at varying actuation forces and distances from nozzle tip, developed flow only, from the Malvern Spraytec.}
\label{fig:SMDbPlot}
\end{figure}
\newpage

\newpage
\newpage
\begin{figure}[h]
\centering\includegraphics[width=0.9\linewidth]{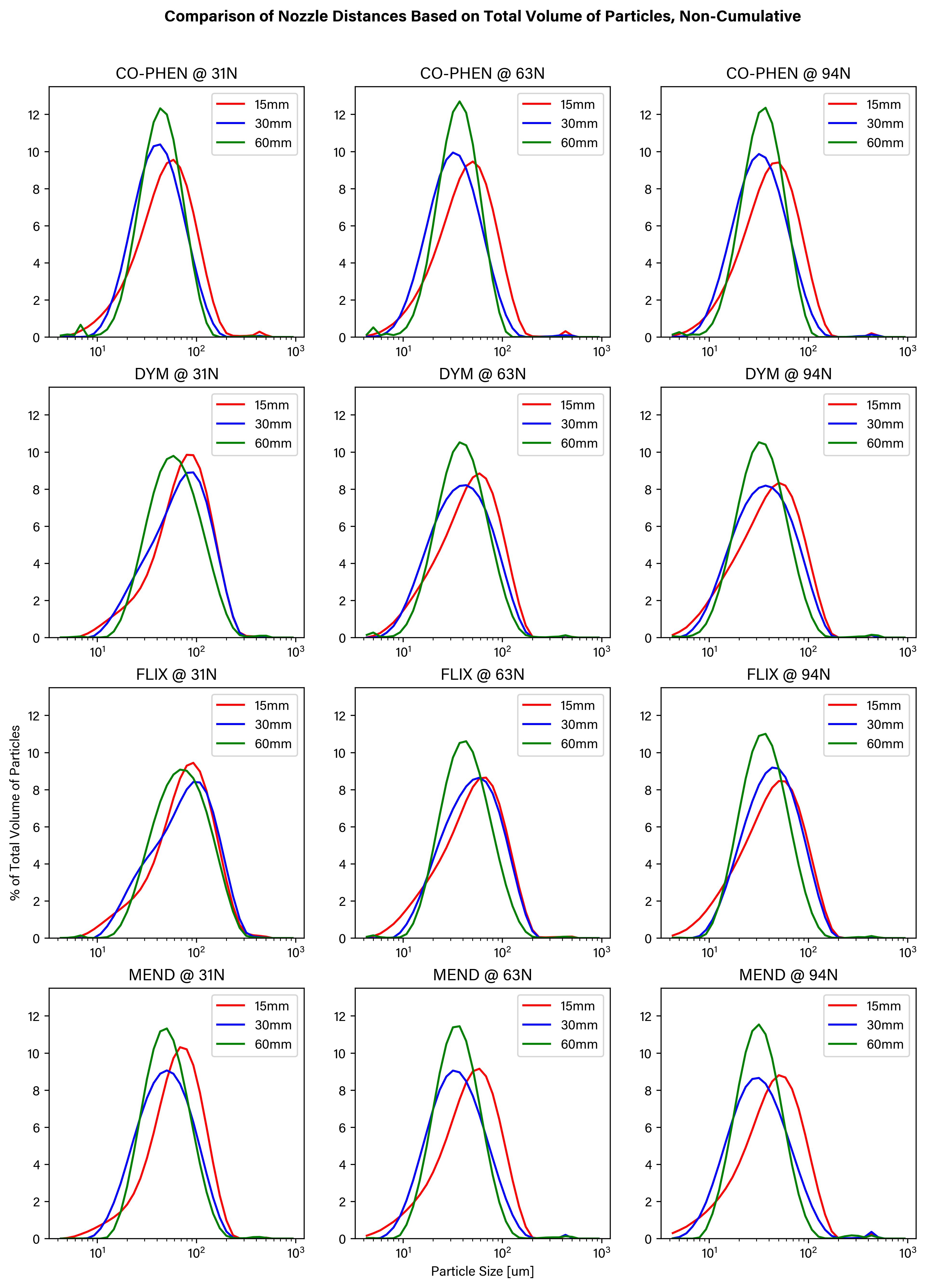}
\caption{Non-cumulative droplet size distribution comparing samples at varying distances to the nozzle tip.}\label{fig:distDSD-nonCum}
\end{figure}
\newpage
\begin{figure}[h]
\centering\includegraphics[width=0.9\linewidth]{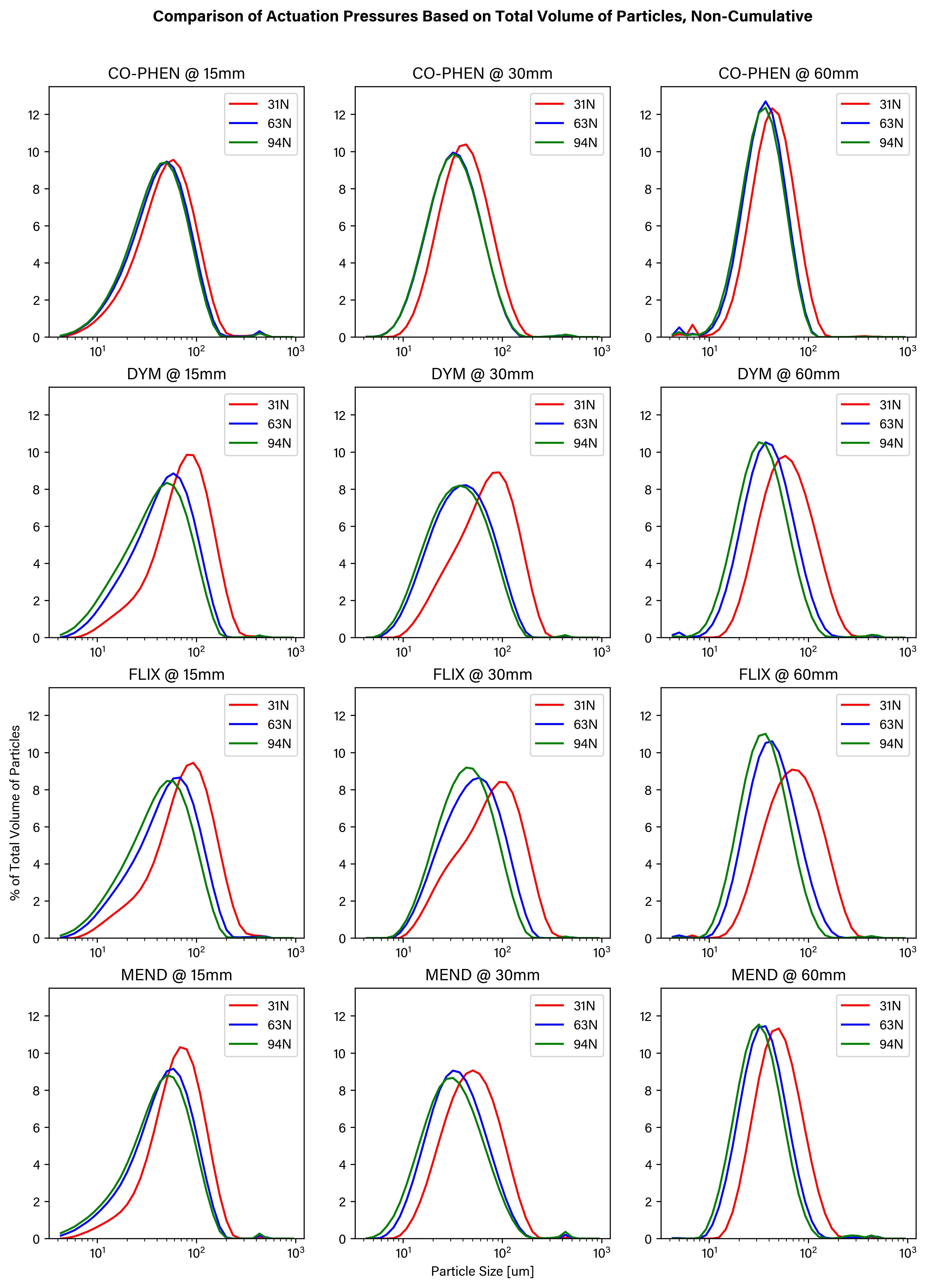}
\caption{Non-cumulative droplet size distribution comparing actuation forces. }\label{fig:PressDSD-nonCum}
\end{figure}
\newpage

\end{document}